\documentclass{article}
\usepackage{longtable}
\usepackage{booktabs,tabularx,siunitx}
\usepackage[numbers,sort&compress,round]{natbib}
\usepackage{tikz}
 \usetikzlibrary{arrows.meta,positioning,shapes.geometric,fit}
\usepackage{PRIMEarxiv1}
\usepackage{natbib} 
\usepackage{tikz}
\usetikzlibrary{arrows.meta,positioning,shapes.geometric}
\usepackage[utf8]{inputenc} 
\usepackage[T1]{fontenc}    
\usepackage{hyperref}       
\usepackage{url}            
\usepackage{booktabs}       
\usepackage{amsmath}        
\usepackage{amsfonts}       
\usepackage{amssymb}        
\usepackage{nicefrac}       
\usepackage{microtype}      
\usepackage{lipsum}
\usepackage{fancyhdr}       
\usepackage{graphicx}       
\graphicspath{{media/}}     
\usepackage{enumitem}       

\title{Extension of the Fusion Power Plant Costing Standard}

\author{
  Simon Woodruff, Alicia Durham, Alex Higginbottom, Chris Raastad \\
  Woodruff Scientific Ltd \\
  1 Kings Meadow \\
  Oxford\\
  \texttt{\{Simon Woodruff\}simon@woodruffscientific.com} \\
}

\begin{document}
\maketitle

\begin{abstract}
This paper documents the work of the Clean Air Task Force (CATF) International Working Group (IWG) on Fusion Cost
Analysis in 2024--2025, and the methodological extensions implemented in the CATF-supported branch of the
pyFECONs fusion power-plant costing framework. Using the standards-aligned chart-of-accounts and physics-to-economics workflow established by ARPA-E \cite{Woodruff2026CostingFramework}.  The IWG
development reorganizes and deepens the framework around three architecture-defining cost-driver tracks:
\emph{Magnetic Fusion Energy (MFE)}, \emph{Inertial Fusion Energy (IFE)}, and \emph{Magneto-Inertial Fusion Energy
(MIFE)}. In particular, the generic driver placeholder in Account~22.1.3 (``Coils or Lasers or Pulsed Power'') is treated
as a controlled ``swap-point'' and replaced by a full cost-account development for the dominant driver in each class:
magnets (TF/PF/CS) for MFE, lasers/driver modules for IFE, and pulsed-power systems for MIFE, enabling auditable
traceability from requirements and geometry to rolled-up plant costs.  On top of this driver-centric foundation, we introduce a probabilistic costing layer that compounds materials price
uncertainty, TRL-based maturity uncertainty, and learning-curve uncertainty into cost distributions, and we outline a
planned alignment of reported uncertainty bands with the AACE estimate-class accuracy ranges (``cone of uncertainty'')
to support consistent communication of estimate maturity. We then describe safety-informed costing that enumerates
fusion-relevant hazards and maps mitigating systems, structures, and provisions into standardized accounts, together
with scenario-parameterized regulatory and financial adders (licensing and insurance). Finally, we document expanded
macroeconomic and finance parameterization and a value-metrics module that complements LCOE with investment and
planning measures (NPV, IRR/MIRR, revenue requirements, WACC-based annualization, and residual/follow-on value),
all computed from the same COA-mapped outputs. Collectively, these additions convert a deterministic, standards-aligned
costing backbone into an extensible analysis environment suitable for transparent sensitivity studies, uncertainty
propagation, and safety- and finance-coupled interpretation of fusion pilot-plant and NOAK scenarios.
\end{abstract}

\section{Introduction}
\label{sec:introduction}

Economic credibility has become a first-order requirement for fusion energy. As concepts mature from physics demonstrations toward integrated pilot plants and early commercial facilities, stakeholders increasingly require cost estimates that are (i) transparent and auditable, (ii) comparable across diverse fusion architectures, and (iii) traceable to explicit technical assumptions rather than embedded in opaque scaling relations or single-point levelized outputs. These needs motivated the standards-aligned fusion costing workflow summarized in the ARPA-E support effort (2017--2024), which evolved from ARIES-style capital-cost studies into an auditable, chart-of-accounts (COA) framework implemented in FECONs and released as the open-source Python codebase pyFECONs. 

This paper documents the subsequent development performed under the Clean Air Task Force (CATF) International Working Group (IWG) on Fusion Cost Analysis in 2024--2025, and consolidates the methodological extensions implemented in the CATF-supported branch of the pyFECONs framework. The baseline COA structure and physics-to-economics pipeline established in \cite{Woodruff2026CostingFramework} is adopted here as the primary reference and as the stable interface for comparability: power balance and net electric output fix the normalization basis; engineering-constrained geometry and subsystem sizing produce auditable quantities; subsystem estimates map into standardized accounts; and rolled-up capital and annualized costs yield LCOE as a baseline comparability metric. The CATF IWG effort extends that baseline in ways aimed at improving policy relevance, licensing relevance, and decision usefulness while preserving the same standards-aligned account mapping and report traceability.

\subsection{CATF IWG objectives and development approach}
\label{sec:intro_objectives}

The CATF IWG operated with quarterly plenary meetings and (for active contributors) a regular weekly implementation cadence. The core objective was not to replace the standards-aligned costing backbone, but to \emph{add controlled analytical capabilities} that enable disciplined sensitivity studies over uncertainty, safety posture, regulatory/insurance assumptions, financing structure, and value metrics. In practice, the development emphasized:
\begin{enumerate}
  \item \textbf{Reproducible workflows and usability:} a live Colab\,+\,Overleaf execution and report-generation workflow, plus a Python-library/API pathway enabling web-facing use and controlled report export;
  \item \textbf{Modular extensions beyond baseline LCOE:} finance/economic valuation measures and risk/uncertainty features implemented as optional layers that consume the same COA-mapped outputs;
  \item \textbf{Fusion-specific cost-driver deepening:} targeted bottom-up cost bases and benchmarks (e.g., power supplies, pulsed power components, direct energy conversion) consistent with the standardized account structure;
  \item \textbf{Safety- and regulation-informed costing:} scoping-level safety methodology that makes hazards explicit, translates them into mitigating plant provisions, and maps them into the appropriate direct, supplementary, and financial accounts.
\end{enumerate}

\subsection{Contributions and scope of this paper}
\label{sec:intro_contributions}

Accordingly, this paper makes four principal contributions.
First, it provides a consolidated record of the CATF IWG program of work (topics, cadence, and documented milestones), including a chronology of meeting topics and a list of documented contributors, establishing provenance for the methodological evolution.
Second, it documents the \emph{implementation mechanisms} used to preserve auditability---notably template-driven \LaTeX{} report generation directly from computed quantities and standardized account rollups.
Third, it presents the major \emph{methodology extensions} developed in the CATF IWG branch:
(i) probabilistic costing that compounds materials uncertainty, TRL-based maturity uncertainty, and learning-curve uncertainty into cost distributions;
(ii) safety-informed costing and explicit mapping of safety-driven scope into standardized accounts (including links to licensing fees and insurance adders);
(iii) macroeconomic and finance parameterization (inflation, escalation, interest/discount factors) consistent with account rollups; and
(iv) an expanded economics-measures module that complements LCOE with investment/value metrics such as NPV, IRR/MIRR, revenue requirements, WACC-based annualization, and residual/follow-on value (including an APV framing).
Fourth, it positions these additions as an extensible analysis environment suitable for transparent sensitivity studies and policy-relevant interpretation of pilot-plant and NOAK scenarios, without destabilizing the standards-aligned comparability layer established in \cite{Woodruff2026CostingFramework}.

\subsection{Paper organization}
\label{sec:intro_org}

The remainder of this paper is organized as follows. Section~2 surveys the state of the art in fusion power-plant cost analysis, emphasizing the complementary roles of standardized cost-account frameworks and systems-code-driven design-and-cost workflows \cite{Woodruff2026CostingFramework}. Section~3 describes the baseline ARPA-E standards-aligned costing methodology adopted as the reference implementation for this work \cite{Woodruff2026CostingFramework}. Section~4 then summarizes the CATF extensions to the standard, highlighting the driver swap-point (Account~22.1.3), the coupled electrical infrastructure refinement (Account~22.1.7), and the propagation of safety-informed cost bases through the COA \cite{CATF_IWG_pyFECONs_Extensions_ColabLog,CATF_FusionCostModelPortal}. Section~5 documents the 2024--2025 chronology of IWG activities, and Section~6 lists participants and contributors with documented discussion inputs. Section~7 details the methodology extensions implemented in the CATF IWG Python framework, including reproducible reporting, probabilistic costing, learning treatment, safety-to-account mapping with licensing and insurance adders, and the expanded economics/value module. Section~8 describes worked examples and the open Colab workflow used to reproduce results \cite{CATF_IWG_pyFECONs_Extensions_ColabLog}. Section~9 discusses the relationship of this approach to prior fusion and power-plant costing traditions, Section~10 outlines further work directions, and Section~11 concludes with a summary of the capability and its intended use in credible, comparable fusion techno-economic analysis.

\section{State of the Art in Power Plant Cost Analysis}

Fusion power--plant cost analysis is currently shaped by two major methodological trends. The first is the use of \emph{standardised cost account frameworks}---a concept--agnostic taxonomy for organising plant cost elements that enables auditable, cross--concept comparisons. Such account structures were introduced in international fission and nuclear project practice (e.g.\ IAEA--style accounts) and have been adapted for fusion in programmes such as ARPA--E tool development and related benchmarking efforts \cite{arpae_fecon}. The second trend is the \emph{systems--code approach}, in which a reactor concept is defined by enforcing plasma physics and engineering design constraints and then attaching cost correlations or bottom--up modules to the resulting, self--consistent design point(s) \cite{Kovari2014,Kovari2016,PROCESSgit,EUROfusion2016}. 

These two strands interact but should be distinguished in the literature. Systems codes typically become increasingly elaborate and concept--specific because they embed physics and engineering constraints that differ fundamentally across configurations. By contrast, standard cost accounts are intentionally stable across concepts, serving as a common language even as the underlying design and subsystem models evolve. Reflecting this structure, the state of the art can be grouped into: (i) account--based costing frameworks and their extensions to FOAK risk, learning, and system value; and (ii) systems codes used to generate constrained reactor designs and associated cost outputs.

\subsection{Account--Based Costing Frameworks and Extensions}

\subsubsection{ARPA--E and US Standardised Frameworks (FECONs / pyFECONs)}
A major advance in US practice is the development of modular, account--based costing frameworks under ARPA--E programmes, exemplified by the FECONs/pyFECONs tools \cite{arpae_fecon}. These tools codify an evolution from ARIES--style empirical correlations toward bottom--up subsystem models calibrated to vendor and EPC data, with a view to enabling auditable and cross--concept comparisons within a stable account structure. For example, the magnet system is decomposed into manufacturing and material cost components (superconductor, copper, structural steel, insulation, support structure), with cost scaling relationships fitted to a dataset of reference conductor designs; similar decompositions exist for blankets, vacuum systems, tritium handling, and balance--of--plant components \cite{arpae_fecon}.

Applied across multiple ARPA--E funded concepts (compact tokamaks, stellarators, magneto--inertial fusion, and alternative geometries), this common account structure has supported identification of robust cost drivers such as the magnet system, tritium plant, and thermal power conversion train, and has enabled comparisons of fusion against other generation technologies \cite{arpae_fecon}. In this sense, the ARPA--E framework is emerging as a de facto reference point in US public fusion programmes, intended to support consistent benchmarking across concepts and against non--fusion options \cite{arpae_fecon}.

\subsubsection{Account--Based Bottom--Up Models in DEMO/ARIES--Type Studies}
Public DEMO--oriented studies continue to provide the backbone of detailed capital cost modelling for tokamak--based fusion power plants \cite{wenninger2017,aries_comparative}. In Europe and Korea, PEC--style approaches and related workflows link plasma physics choices, engineering parameters, and technology options to total overnight capital cost and cost of electricity, enabling parametric studies over plant size, wall loading, power density, and availability \cite{wenninger2017,aries_comparative}. Earlier ARIES studies and their successors formulate cost--of--electricity optimisation problems over design space, showing, for example, that higher power density and more compact designs can move fusion towards cost ranges competitive with advanced fission under idealised assumptions about technology readiness and availability \cite{aries_comparative,mitnuclear2018}.

Within these studies, major plant subsystems (magnets, vacuum vessel, blanket and shield, tritium systems, power conversion, buildings and balance of plant) are commonly represented via parametric cost correlations, often derived from fission experience and adapted to fusion geometries and material inventories \cite{aries_comparative,wenninger2017}. This modelling layer remains central for design trade--offs and sensitivity analysis, but on its own tends to under--represent first--of--a--kind (FOAK) construction and project delivery risk, motivating more recent integration with megaproject evidence \cite{mitnuclear2018,step_rcf}.

\subsubsection{FOAK Evidence and Structural Uncertainties}
Beyond STEP, some of the most instructive fusion--specific cost evidence comes from FOAK experimental devices. The National Compact Stellarator Experiment (NCSX) in the United States was cancelled after substantial cost escalation relative to early, externally validated estimates, with post--mortem analyses highlighting severe underestimation of manufacturing complexity, rework, and quality assurance effort for novel components \cite{neilson2010}. Together with experience from ITER and other large--scale scientific facilities, this has reinforced the view that FOAK fusion programmes are particularly exposed to optimism bias in manufacturing, integration, and regulatory interfaces, not just in headline capital cost categories \cite{step_rcf,neilson2010}.

In the United States, compact pilot plant cost--driver studies similarly emphasise sensitivity to assumptions about availability factors, component lifetimes, replacement intervals, and learning rates, but these analyses often remain closer to engineering cost models, with less explicit integration of RCF--style megaproject corrections than is now seen in UK practice \cite{wade2021,mit_fusion_value}. Across both UK and US work, a persistent issue is that the most informative data on cost escalation are either incomplete or proprietary: public data often cover only final investment decision and final outturn costs, while early estimate histories and detailed risk registers remain inaccessible \cite{step_rcf,arpae_fecon}. This scarcity constrains the construction of fusion--specific reference classes and forces reliance on proxy domains for statistical correction.

\subsubsection{Megaproject Risk, Optimism Bias, and Reference Class Forecasting}
A key contemporary shift, strongly influenced by UK work on the STEP programme, is the explicit treatment of megaproject risk, optimism bias, and cost overrun statistics on top of engineering cost models \cite{step_rcf,mitnuclear2018}. Fusion power plants are now explicitly framed as megaprojects: large, complex, multi--stakeholder ventures with budgets often exceeding \$1~billion, long construction times, and strong interdependencies among subsystems and stakeholders \cite{step_rcf}. Empirical analyses of large industrial and nuclear projects show that early cost estimates are systematically poor predictors of final outturn cost, with both mean cost and schedule overruns increasing as design and construction progress \cite{mitnuclear2018,step_rcf}. To mitigate this, leading institutions have begun to import reference class forecasting (RCF) into fusion cost practice, regressing ``inside view'' estimates toward empirical reference--class distributions and applying uplifts to target specified overrun probabilities (e.g.\ P50, P70, P80) consistent with guidance such as the UK Treasury Green Book and the Infrastructure and Projects Authority (IPA) \cite{hmt_greenbook,ipa_rcf}.

For fusion, the central methodological challenge is the lack of a historical fleet of power plants from which to draw a strictly fusion--specific reference class. Programmes such as STEP therefore assemble mixed reference classes drawn from sectors with similar novelty and complexity, including fission, large scientific facilities (ITER, CERN, ESS), space missions, and major transport infrastructure \cite{step_rcf}. This raises questions about transferability of risk patterns across domains, but is currently the only pragmatic route to empirically grounded optimism--bias corrections in early--stage fusion plant estimates \cite{mitnuclear2018,step_rcf}.

\subsubsection{System--Level Value, Target Costs, and Private--Sector Perspectives}
A further account--based strand evaluates fusion not only at the plant level but in the context of whole energy systems. Recent MIT studies embed fusion plants in integrated energy--system models and derive economic value and optimal deployment as a function of overnight capital cost, operating characteristics, and policy constraints \cite{mitnuclear2018}. These analyses show that under stringent climate policy, fusion can capture a large share of global electricity supply and substantially reduce decarbonisation costs if capital costs fall to a few thousand US dollars per kilowatt and availability is high, whereas at higher costs its optimal share shrinks dramatically \cite{mit_fusion_value}.

In parallel, industry and investor reports synthesise private--sector views on levelised cost of energy (LCOE) and learning curves for commercial fusion plants \cite{euvc_fusion,fia_report2023}. Such work typically argues that FOAK commercial plants in the 2030s are more likely to exhibit LCOE in the 150--200~\$/MWh range, with 60--100~\$/MWh figures only plausible for NOAK plants after substantial learning and scaling \cite{euvc_fusion}. These studies highlight enabling technologies such as high--temperature superconductors as potential sources of cost decline, given historically high learning rates for analogous advanced materials and manufacturing chains \cite{euvc_fusion,fia_report2023}.

\subsection{Systems--Code--Driven Design--and--Cost Workflows}

During conceptual design, a number of 0D/1D systems codes generate self--consistent fusion plant designs under physics, technology and engineering constraints and then attach costing models to evaluate figures of merit such as cost of electricity \cite{Najmabadi2010,Dragojlovic2010,EUROfusion2016}. The ARIES systems code, developed for a series of advanced tokamak studies, provides modular plasma, power--core and costing models and has recently been re--implemented with updated physics, object--oriented structure and enhanced optimisation and visualisation capabilities to examine a broad trade space rather than sensitivities around a single baseline design \cite{Dragojlovic2010,Najmabadi2010}. EUROfusion and associated groups employ in--house systems tools of similar scope for EU DEMO and related concepts, typically coupling 0D plasma models with simplified engineering and costing to generate consistent design points that are then passed to higher--fidelity physics and engineering analyses \cite{EUROfusion2016}. In parallel, various national laboratories and universities maintain more specialised 0D plant simulators focused on particular aspects such as transient power and particle balance or technology choices, which play a similar role in mapping viable design regions prior to detailed design \cite{EUROfusion2016}.

PROCESS is UKAEA's long--standing 0D/1D fusion power plant systems code, used to generate self--consistent plant designs under physics and engineering constraints and to optimise figures of merit such as cost of electricity; it is now being modernised and wrapped in Python for integration into wider digital design workflows \cite{Kovari2014,Kovari2016,PROCESSgit}. Recent work increasingly employs PROCESS as a fast whole--plant integrator embedded in multi--tool workflows for STEP/DEMO--like studies, handing baseline design points and requirements to higher--fidelity transport, equilibrium and engineering tools \cite{DigitalFusion2024}. BLUEMIRA, developed by UKAEA as a Python--based, modular, multi--fidelity fusion reactor design framework, aims to reduce tokamak conceptual design times from months to minutes by unifying equilibrium, magnetostatics, 3D geometry/CAD, neutronics--model generation, and simplified power and fuel--cycle models in scripted, reproducible workflows \cite{BluemiraDocs,UKAEABluemira}. It provides explicit roles for modellers, reactor designers and developers, includes an interface to the PROCESS systems code and other physics solvers, and is being deployed as an integrated environment for STEP and DEMO design studies, in which PROCESS supplies rapid plant--level consistency while BLUEMIRA handles geometry--resolved, constraint--rich optimisation of coils, divertor and other subsystems \cite{BluemiraDocs,Graham2025}.

\subsection{Cross--Cutting Themes and Open Problems}

Across leading UK and US institutions, several cross--cutting themes define the state of the art. First, there is a clear recognition of the FOAK versus NOAK gap: FOAK plants are expected to be much more expensive than long--run targets, but quantitative learning--curve models for fusion remain poorly constrained due to the absence of deployment data and uncertainty about ultimate market size \cite{step_rcf,euvc_fusion}. Second, data scarcity and confidentiality remain major obstacles, since many informative benchmarks for escalation and risk (ITER, NCSX, consultant databases, EPC cost books) are not fully transparent \cite{step_rcf,arpae_fecon,neilson2010}. Third, account--based frameworks and systems--code workflows must continue to evolve to provide fair comparisons across diverse fusion concepts, requiring a common taxonomy that can accommodate divergent physics, plant architectures, and product offerings (grid power, process heat, hybrid systems) \cite{arpae_fecon,aries_comparative}.

Finally, integration of regulatory, financing, and delivery--model effects into fusion cost practice is an emerging frontier. Recent system--level and commercial analyses increasingly treat regulation, financing structure, and project delivery models as first--order cost drivers rather than exogenous constraints, but rigorous quantitative integration of these factors into fusion--specific tools remains at an early stage \cite{euvc_fusion,step_rcf}. Taken together, these developments indicate that fusion cost analysis is moving from isolated concept--specific estimates toward unified, auditable approaches that (i) resolve subsystem costs in detail within stable accounts, (ii) apply empirically grounded megaproject uplifts, and (iii) assess economic value in whole--system contexts, while grappling with severe data limitations and FOAK uncertainty \cite{step_rcf,arpae_fecon}.

\section{Methodology}
\label{sec:background}

This paper builds directly on the costing methodology developed over the period 2017-2024 by ARPA-E and consolidated in \cite{Woodruff2026CostingFramework}, which we adopt as the \emph{primary reference} for (i) the standards-aligned chart-of-accounts (COA) structure, (ii) the physics-to-economics workflow, and (iii) the implementation philosophy of traceable, auditable plant costing for fusion power plants. The present work extends that foundation by documenting the additional capabilities developed within the Clean Air Task Force (CATF) International Working Group (IWG) on Fusion Cost Analysis.

\subsection{Motivation: credible, comparable fusion TEA}
As fusion concepts move from physics demonstrations toward integrated pilot plants, economic credibility becomes a first-order requirement. The core challenge is to produce cost estimates that are transparent, comparable across architectures, and traceable to underlying technical assumptions rather than embedded in opaque scalings or single-point LCOE outputs. Achieving this requires (a) a consistent accounting taxonomy, (b) explicit translation from physics and engineering constraints to plant equipment and buildings, and (c) disciplined reporting of assumptions and cost bases.

\subsection{Methodological lineage and evolution}
The ARPA-E methodology evolved from ARIES-style systems studies (physics-informed power balance and engineering-constrained radial builds) into a programmatic, portfolio-capable costing workflow. Early ARPA-E work emphasized capital cost estimation and calibration against engineering, procurement, and construction (EPC) intuition through a pilot benchmarking effort with Bechtel and Decysive Systems. Subsequent phases deepened the treatment of balance-of-plant (BOP) and (especially) indirect costs, reflecting the recognition that non-fusion-island scope (layout, buildings, execution methods, construction management, and cost-of-money) can dominate total plant costs. Portfolio-scale application across diverse magnetic, inertial, and magneto-inertial architectures further motivated harmonized assumptions, improved fuel-cycle realism (including tritium handling), and a stronger emphasis on auditability and cross-concept comparability.\\

A central contribution of \cite{Woodruff2026CostingFramework} is the refactor of the costing structure to align with the IAEA / GIF-EMWG / EPRI COA lineage, separating direct costs (pre-construction and construction) from capitalized indirect services, owner’s costs, supplementary costs, and financial costs. This standards-aligned COA is treated as a durable interface between evolving fusion subsystem models and externally interpretable economic outputs. Importantly, fusion-unique systems (heat island, fuel cycle, remote handling, and major replacement schedules) are mapped explicitly into the standardized accounts to preserve comparability while making fusion-specific departures visible and auditable.

\subsection{Physics-to-economics workflow and tool implementation}
The reference workflow follows a sequential pipeline:
(1) establish a physics-informed power balance to fix gross and net electric output (the normalization basis for $/kWe$ and annual generation);
(2) construct an engineering-constrained radial build to determine principal geometry of the heat island (and quantities that drive component sizing);
(3) size dominant cost-driver systems (e.g., magnets, lasers, pulsed power, power supplies) in a manner coupled to the plant operating point;
(4) translate thermal performance and layout into BOP equipment and buildings, anchored where appropriate to transparent external baselines;
(5) assemble overnight construction cost from explicit equipment and structures; then
(6) roll up indirect, owner, supplementary, and financial accounts using well-defined accounting conventions, followed by annualized costs and LCOE.\\

This workflow was implemented first in the spreadsheet-based Fusion Economics code (FECONs) and then released as an open-source Python implementation (pyFECONs) emphasizing transparency, traceability, and extensibility. In the same reference, the post-release evolution supported by CATF is identified as a closed-source branch with a web interface and expanded costing features, including expanded treatment of safety systems. The present paper documents that CATF IWG-driven evolution and the methodological extensions developed to support policy-relevant and licensing-relevant costing questions.

\subsection{Modular extensions beyond baseline LCOE}
An architectural choice is made that is particularly important for the CATF IWG work: maintain a stable, standards-aligned baseline implementation for reproducibility and like-for-like comparisons, while implementing higher-fidelity or decision-context-specific analyses as optional modules. Examples include design-for-maintainability and RAMI cost-out modules, retrofit/repowering scenarios, materials scarcity and manufacturability risk, and extended financial valuation metrics (e.g., NPV, IRR/MIRR, payback, revenue requirements, and WACC-based analyses) \cite{Woodruff2026CostingFramework}. This modular structure provides a disciplined mechanism to incorporate CATF priorities (notably safety- and regulation-informed costing) without destabilizing the reference COA-mapped workflow that enables comparability across concepts and studies.
\begin{figure}[htbp]
\centering
\begin{tikzpicture}[
  font=\small,
  node distance=9mm and 10mm,
  box/.style={draw, rounded corners, align=center, inner sep=6pt, text width=0.3\textwidth},
  bigbox/.style={draw, rounded corners, align=center, inner sep=8pt, text width=0.5\textwidth},
  arrow/.style={-{Stealth[length=2.2mm]}, thick},
  group/.style={draw, dashed, rounded corners, inner sep=8pt}
]

\node[bigbox] (arpa) {\textbf{ARPA-E Costing Standard (retained)}\\
Standards-aligned COA + physics-to-economics workflow\\
Consistent accounting container for cross-architecture comparability};

\node[box, below=of arpa, xshift=-0.2\textwidth] (ref2213) {\textbf{CATF refinement: Account 22.1.3}\\
Driver ``swap-point'' refined within standard COA\\
\emph{MFE: magnets (TF/PF/CS)}\\
\emph{IFE: lasers/driver modules}\\
\emph{MIFE: pulsed power driver}};
\node[box, below=of arpa, xshift=+0.2\textwidth] (ref2217) {\textbf{CATF refinement: Account 22.1.7}\\
Power supplies / pulse-forming infrastructure\\
Installed cost bases (e.g., \$/J, \$/kW)\\
Module count, lifetime, replacement/derating logic};

\node[bigbox, below=of ref2213, xshift=+0.2\textwidth, yshift=-1mm] (safety) {\textbf{Safety-informed costing (CATF extension)}\\
Safety basis $\rightarrow$ mitigating SSCs and provisions $\rightarrow$ mapped into affected COA accounts\\
(licensing, shielding, confinement, monitoring, emergency systems, etc.)};

\node[group, fit=(ref2213)(ref2217)(safety), label={[align=center]above:\textbf{CATF methodological extensions within the ARPA-E COA}}] (catfgroup) {};

\node[bigbox, below=of safety] (caplcoe) {\textbf{Model outputs (from COA roll-up)}\\
Total Capital Cost (CAPEX) + cost breakdown by account\\
Operating costs and availability assumptions\\
\textbf{Levelized Cost of Electricity (LCOE)}};

\node[box, below=of caplcoe, xshift=-0.21\textwidth] (econext) {\textbf{Economics modeling extension}\\
NPV, IRR/MIRR, payback, revenue requirements\\
WACC-based annualization, financing scenarios\\
Sensitivity/uncertainty propagation};
\node[box, below=of caplcoe, xshift=+0.21\textwidth] (depext) {\textbf{Deployment modeling extension}\\
Build-rate / market-capture constraints\\
Fleet evolution, diffusion limits, learning-by-doing coupling\\
Capacity trajectories and timing scenarios};

\draw[arrow] (arpa) -- (ref2213);
\draw[arrow] (arpa) -- (ref2217);

\draw[arrow] (ref2213.south) |- (safety.north);
\draw[arrow] (ref2217.south) |- (safety.north);

\draw[arrow] (safety) -- (caplcoe);

\draw[arrow] (caplcoe) -- (econext);
\draw[arrow] (caplcoe) -- (depext);


\end{tikzpicture}
\caption{Workflow diagram: the ARPA-E costing standard (COA and workflow) is retained, while CATF refines key driver and electrical
accounts (22.1.3 and 22.1.7) and propagates safety-driven cost bases into other affected accounts. COA roll-up produces capital costs
and LCOE, which then feed the economics and deployment modeling extensions.}
\label{fig:catf_arpae_workflow}
\end{figure}
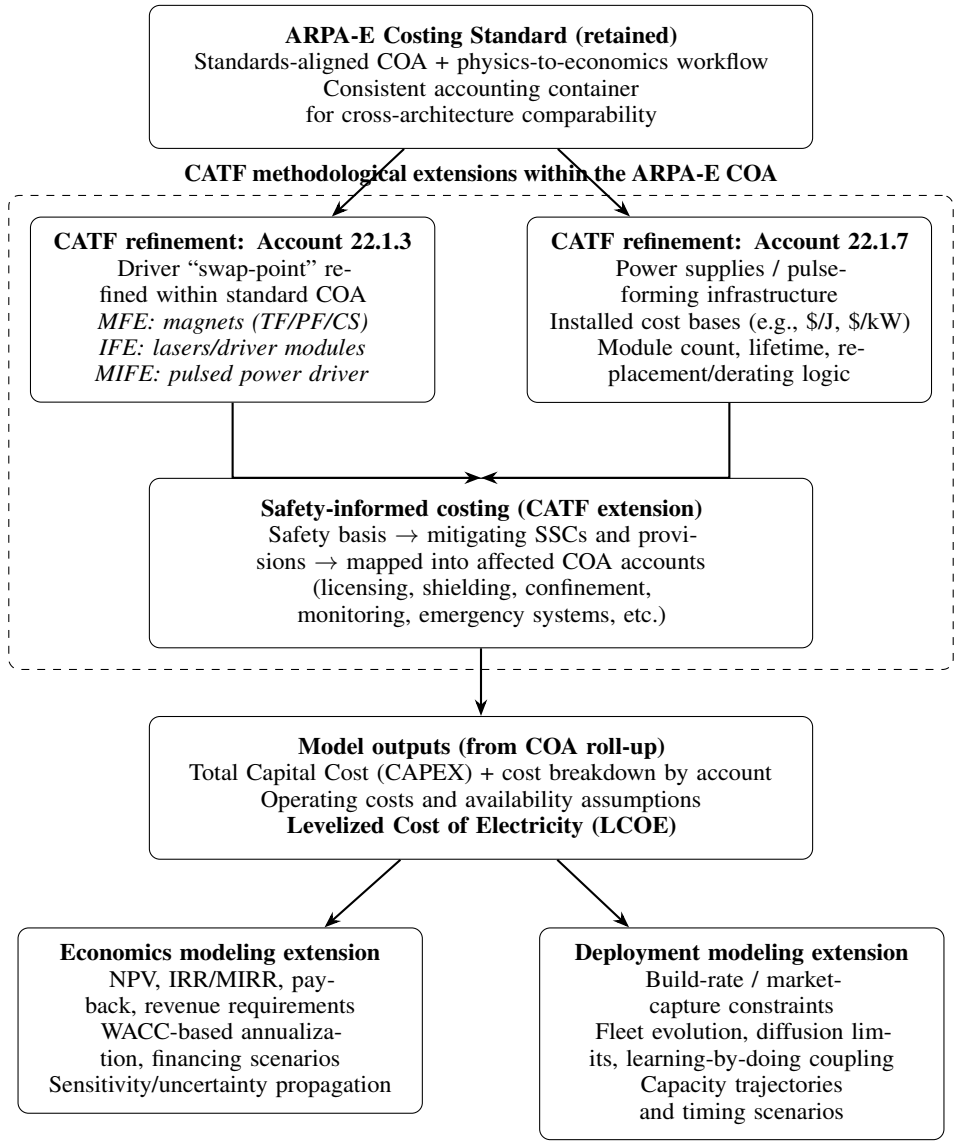

\section{CATF Extensions to the Standard}
\label{sec:iwg_intended_scope}

Figure~\ref{fig:catf_arpae_workflow} summarizes how the CATF IWG extended the ARPA-E fusion costing capability while retaining the
underlying standards-aligned chart-of-accounts (COA) and physics-to-economics workflow as the stable baseline for comparability.
The CATF developments concentrate new methodological fidelity where fusion cost uncertainty is highest: refining the dominant
driver account (22.1.3) and the electrical/power-conditioning account (22.1.7) with bottom-up cost bases and explicit scaling
assumptions, while also propagating safety-informed requirements into the affected COA accounts through mapped mitigating
systems and provisions. These refinements preserve a consistent COA roll-up to total capital costs and LCOE, and then use those
same outputs as the common quantitative foundation for downstream extensions in economics modeling (e.g., revenue requirements,
NPV/IRR) and deployment modeling (e.g., build-rate constraints and fleet evolution), ensuring that higher-level policy and investment
metrics remain traceable to auditable engineering and cost-account assumptions.  The following expands on the major developments from the CATF IWG on Fusion Cost Analysis.

\begin{itemize}

  \item \textbf{Tooling, release, and usability}
  \begin{itemize}
    \item Live, reproducible workflow using Google Colab + Overleaf for code execution and report generation.
    \item Development of a fusion costing \textit{Python library} and use as an \textit{API} for a web-based costing application.
    \item Platform updates spanning: Python Colab notebooks, a packaged Python distribution, and a CATF-released web interface connected to the Python package.
    \item Report-building workflow where CATF-developed additions are maintained as appendices and/or merged into standard reports as explicit cost bases (with the Python script overwriting placeholder variables in the \LaTeX{} report templates).

  \end{itemize}
    \item \textbf{Probabilistic costing and uncertainty propagation}
  \begin{itemize}
    \item Probabilistic approach combining: TRL-driven uncertainty, material-cost uncertainty, experience-curve / learning-curve effects, and resulting cost uncertainty.
    \item Manufacturing cost model to (i) accept a user-specified manufacturing technique or (ii) infer plausible methods from part complexity/size/material applicability, yielding a cost distribution and (optionally) a recommended method.
    \item A manufacturing-method dictionary (e.g., injection molding, 3D printing, CNC machining, stereolithography/SLA) with method viability filtering and comparison logic.
  \end{itemize}

 \item \textbf{Driver-centric cost-basis deepening for MFE, IFE, and MIFE}
\begin{itemize}
  \item \textbf{MFE (magnet-dominated development):}
  \begin{itemize}
    \item \emph{HTS magnet benchmarking and cost-account development:} bottom-up costing of TF/PF/CS magnet systems, including materials and fabrication drivers, intended to replace the generic driver placeholder in Account~22.1.3 with an explicit magnet cost basis.
    \item \emph{Worked examples for fusion ``heat-island'' components:} first wall, blanket, and shielding calculations used to demonstrate how geometry and performance requirements translate into COA-mapped quantities and costs in the Python/Colab workflow and report outputs.
  \end{itemize}

  \item \textbf{IFE (laser/target-dominated development):}
  \begin{itemize}
    \item \emph{Laser target benchmarking:} explicit target costing and throughput logic to support driver-adjacent recurring costs and to improve traceability of IFE-specific consumables.
    \item \emph{Driver/laser account refinement:} development of cost bases and scaling logic for laser/driver modules (mapped to the Account~22.1.3 swap-point) with transparent links to repetition rate, delivered energy, and system modularity.
  \end{itemize}

  \item \textbf{MIFE (pulsed-power-dominated development):}
  \begin{itemize}
    \item \emph{Pulsed-power costing development:} explicit bottom-up treatment of pulsed-power hardware and infrastructure as a first-order cost driver, using installed \$/J cost bases and module replication logic.
    \item \emph{Power supplies as a major cost category:} detailed treatment of supporting electrical plant and pulse-forming infrastructure consistent with Account~22.1.7 framing, preserving separation between the driver (22.1.3) and supporting power-conditioning systems (22.1.7).
    \item \emph{Vendor/supply-chain engagement:} capacitor and switch cost and lifetime inputs, including derating practices and lifetime differences across pulsed-fusion schemes, enabling explicit replacement vs life-of-plant modeling.
  \end{itemize}

  \item \textbf{Cross-cutting (applicable to multiple architectures):}
  \begin{itemize}
    \item \emph{Direct Energy Convertors (DEC):} inclusion and costing treatment development for architectures where direct conversion is material, with COA mapping designed to preserve cross-architecture comparability.
  \end{itemize}
\end{itemize}

  \item \textbf{Safety integration into costing}
  \begin{itemize}
    \item Integration of fusion safety information into the CATF costing code (safety-informed additions/adjustments to relevant cost accounts and supporting methodology).
  \end{itemize}

  \item \textbf{Macroeconomic and finance parameterization}
  \begin{itemize}
    \item Inflation, escalation, interest rates, and discount-factor handling (including consistent translation into capitalized financial accounts and valuation calculations).
    \item “Cost vs.\ value” framing: consistent computation and comparison of LCOE, NPV, IRR (and related measures).
    \item Refinements to NPV handling for NOAK framing and (prototyped) variable discount rates by cost category.
  \end{itemize}

  \item \textbf{Extended valuation metrics module (beyond LCOE)}
  \begin{itemize}
    \item Net Present Value (NPV) with explicit discounting and escalation.
    \item Total Life-Cycle Cost (TLCC).
    \item Revenue requirements (required annual revenue stream for cost recovery / return targets).
    \item Annualized value / equivalent annual cost formulations.
    \item Internal Rate of Return (IRR) and Modified Internal Rate of Return (MIRR).
    \item Simple and discounted payback period.
    \item Weighted Average Cost of Capital (WACC) and its use within NPV and revenue-requirement analyses.
  \end{itemize}

\end{itemize}


\section{Chronology of Work: CATF International Working Group on Fusion Cost Analysis (2023--2025)}

The CATF International Working Group (IWG) on Fusion Cost Model Analysis operated with quarterly plenary meetings and (for active contributors) a regular weekly working cadence (Friday calls were used for implementation-team discussions, with additional ad hoc coordination via shared tools such as Google Colab/Overleaf and the PyFECONs codebase). %

\subsection{Chronology of documented topics (meetings and weekly working sessions)}

\begin{longtable}{p{2.8cm} p{3.4cm} p{9.0cm}}
\hline
\textbf{Date} & \textbf{Meeting type} & \textbf{Topics worked on (documented)} \\
\hline
\endfirsthead
\hline
\textbf{Date} & \textbf{Meeting type} & \textbf{Topics worked on (documented)} \\
\hline
\endhead
\hline
\endfoot

19 Jul 2024 &
Quarterly plenary (Kickoff; Boston + Zoom) &
IWG kickoff and orientation; review of fusion costing methodology/history and major cost drivers; demonstrations and capability discussions including: (i) CATF costing code demo via Google Colab/Overleaf, (ii) integration of safety information into the CATF code, (iii) development of a fusion-costing Python library and its use as an API for a web-based costing application, (iv) maintenance strategy selection to maximize system value, (v) ``value-led'' approaches to fusion costing, (vi) TEA of DT magnetic confinement fusion power plants, and (vii) multi-objective whole-plant optimization workflows. The agenda also included an open discussion with private-sector participants and an implementation-planning working session for the next quarter. %

\\
& Weekly implementation cadence (plan) &
The kickoff minutes specify an expectation of ongoing weekly review/engagement by participants for capability development (weekly commitment noted as a working rule). \\%

16 Aug 2024 &
Weekly working sessions (costing tools / standards) &
Web-based costing calculator demonstrated (signup $\rightarrow$ project creation $\rightarrow$ cost calculation $\rightarrow$ results display $\rightarrow$ PDF export). Discussion focused on: defining the intended audience/standard for the web interface; ensuring correctness of costing calculations and I/O; visualization; and the need for machine-readable outputs/standards. A fixed charge rate calculation spreadsheet was presented and discussed, including treatment of inflation, returns, and constant vs.\ current dollars (with reference to NETL-style results). Discussion also covered prevailing costing/reporting standards and what belongs in a high-level dashboard (e.g., LCOE; power balance; capex/opex breakdowns; fuel costs; construction time; cost-of-electricity breakdowns and visualizations). Action items included: creating a PyFECONs Colab for easier experimentation and machine-readable outputs; adding high-level documentation of calculations, categories, sources and methods; collecting feedback for a results dashboard; and sharing/incorporating the fixed charge spreadsheet into PyFECONs. \\

25 Oct 2024 &
Quarterly plenary (Harwell + Zoom) &
Second plenary agenda emphasized: review of IWG goals/rules and the standard methodology/history/cost drivers; new capabilities demo using Colab/Overleaf; safety integration; Python library/API for a web-based application; workflow discussions (including Google Sheets); benchmark costing topics (laser targets; HTS magnets); continued multi-objective design optimization workflows; plus private-sector commentary and an implementation-planning session for the subsequent quarter. \\

22 Nov 2024 &
Weekly working session (MIF/MagLIF kickoff; planned agenda) &
Kickoff agenda for a Magneto-Inertial Fusion (MIF/MagLIF) costing cycle included: learning curves; capacitor derating for lifetime (trade-offs vs.\ scheduled replacement); supply-chain engagement; and developer questions related to costing tools. \\

29 Nov 2024 (Week 3, MIF) &
Weekly working session &
Topics covered included: MIF power balances; capacitor supply-chain engagement (vendor list engaged); Python package updates; Colab updates including a usage video; CATF MIF report status; what could already be done for fusion pilot plant (FPP) costing; and whether RD\&D costs could be estimated. Discussion also centered on financing frameworks for infrastructure projects (e.g., Regulated Asset Base models as a successor to CfD in the UK) and how to represent financing structures in costing models. \\

06 Dec 2024 (Week 4, MIF) &
Weekly working session &
Topics covered included: first wall / blanket / shield documentation and Python implementation; deployment modeling using S-curves; probability distributions and uncertainty; references supporting contingency justification; and FPP-related matters. \\

13 Dec 2024 (Week 5, MIF) &
Weekly working session &
Agenda/recap emphasized: orientation/purpose for newcomers; recap of first wall/blanket/shield calculations; pulsed-power vendor and supply-chain engagement; updates to PyFECONs platforms (Colab, Python package, web interface); and direct energy convertors. A draft paper on ``Contingency in Fusion Cost Estimation'' was circulated for discussion/reuse. \\

Early Jan 2025 (``last night''; email dated 3 Jan 2025) &
Weekly working session &
Near the end of the MIF cycle: discussion of additional economic metrics beyond LCOE (Net Present Value and Adjusted Present Value); deep-dive into a MIF-relevant cost account (power supplies) including circuit-type cost drivers and deriving \$/J installed costs; capacitor derating for lifetime (adding capacitors to operate at lower voltage as life-of-plant components); and a discussion sparked by commodity price volatility and its implications for cost estimation (including calibration of distributions using historical/trading data). \\

31 Jan 2025 &
Weekly call cancelled; focus on deliverables &
Regular meeting cancelled due to deliverables; priority shifted to releasing the MIF code online through a web portal and completing a MIF concept overview for near-term release. Plan stated to reconvene after Fusion X Invest and then proceed to a magnetic fusion energy (MFE) cycle. \\

28 Feb 2025 -- 21 Mar 2025 &
Release-schedule milestones (email) &
Schedule articulated due to ongoing Magneto-Inertial Fusion Energy (MIFE) work: 28 Feb 2025---``bringing capabilities together'' for a GitHub release (MIFE Colab + Overleaf appendices/cost bases)\\

7 Mar 2025 & Weekly call & release of capabilities and MIFE report to CATF including an analysis for a LANL plasma-jet concept (under review by LANL).\\ 

14 Mar 2025 & Weekly call & final release of MIFE capabilities and a half-day webcast. \\

21 Mar 2025 & Weekly call & MFE cohort start, although support ended. \\

04 Apr 2025 &
Governance/cadence change (email) &
Statement that the team did not expect to return to regular Friday sessions while focusing on website-facing costing code updates; potential reformatting of the interaction cadence into monthly calls. \\

\end{longtable}

\paragraph{Limitations of the chronology.}
Only the topics explicitly captured in the available meeting minutes and email summaries are included above; not every weekly meeting in 2024--2025 has a surviving written summary in the provided corpus. (Where the meeting date is implied rather than explicitly stated, it is marked as inferred.)

\section{Participants and contributors (documented)}

\subsection{Participants with documented discussion contributions (presentations, questions, circulated drafts)}

\begin{itemize}
  \item \textbf{Sehila Gonzalez de Vicente (CATF):} chaired/welcomed and led plenary meeting logistics/rules and working-group framing. %

  \item \textbf{Simon Woodruff (Woodruff Scientific):} introduced methodology/history/cost drivers in plenaries; led/organized weekly implementation-team agendas; coordinated supply-chain engagement (e.g., capacitors) and platform releases (Colab/package/web portal). %

  \item \textbf{Alex Higginbottom (Woodruff Scientific):} demonstrated costing code capabilities (Colab/Overleaf) in plenaries; identified as a weekly meeting lead in Aug 2024 follow-up. %

  \item \textbf{Alicia Durham (Woodruff Scientific):} contributed on integrating safety information into the CATF code; participated in discussion on web tool direction/standards. %

  \item \textbf{Chris Raastad (Woodruff Scientific):} presented the web-based costing calculator and drove discussion on standardization, machine-readable outputs, and dashboard requirements (Aug 2024 summary). %

  \item \textbf{Samuel Ward (Woodruff Scientific):} presented on value-led costing (plenary); presented/discussed financing models (e.g., RAB) and how to capture them in costing; presented NPV/APV as alternative metrics in the MIF weekly cycle. %

  \item \textbf{Layla Araiinejad (MIT):} presented techno-economic analysis of DT magnetic confinement fusion power plants (plenary). %

  \item \textbf{Jacob Schwartz (PPPL):} presented on maintenance strategies to maximize system value (plenary). %

  \item \textbf{Omer Muhammad (nTtau Digital):} presented on use of the costing framework for multi-objective whole power-plant optimization via automated workflows (plenary). %

  \item \textbf{Wayne Meier:} presented and discussed a spreadsheet-based approach to fixed charge rate calculations, with implications for PyFECONs integration. %

  \item \textbf{Jim Gaffney (Focused Energy):} participated in discussion of the web tool, standard framework direction, and fixed charge rate approach. %

  \item \textbf{Maria B. Hancock (Rutherford Energy Venture):} raised a question about commodity price volatility and its role in fusion cost analysis/uncertainty calibration; also referenced in financing context (Rutherford Energy Venture approach). %

  \item \textbf{Geoffrey Rothwell (Independent):} circulated a draft on contingency in fusion cost estimation for group discussion; contingency references were also part of the MIF weekly topics. %

  \item \textbf{Anika Stein:} scheduled/presented benchmark costing of laser targets (Oct 2024 plenary agenda). %

  \item \textbf{Honghai Song (Canyon Magnets):} scheduled/presented benchmark costing of HTS magnets (Oct 2024 plenary agenda). %
\end{itemize}



\section{Methodology Extensions Implemented in the CATF IWG Python Framework}
\label{sec:method_extensions}

The baseline costing workflow (power balance $\rightarrow$ radial build $\rightarrow$ dominant driver sizing $\rightarrow$ BOP translation $\rightarrow$ indirects $\rightarrow$ LCOE reporting) is implemented as a standards-aligned, auditable Python pipeline that computes cost accounts, annualized costs, and then writes results directly into \LaTeX{} report artifacts. In the baseline implementation, the script (i) defines global assumptions and option sets, (ii) computes the power balance, (iii) computes capital cost accounts (Categories 10--60), (iv) computes annualized costs (Categories 70--90), and (v) computes LCOE and writes outputs into \LaTeX{} tables \cite{Woodruff2026CostingFramework}.  Note that all of these extensions are available in an open Google Colab \cite{CATFcolab}, and their implementation as integral parts of the web-based platform is ongoing.
In the CATF IWG effort, we developed (and prototyped in the companion Python notebook used for working sessions) several methodological extensions that sit ``on top of'' the baseline accounting and reporting pathway. These extensions preserve like-for-like comparability of the baseline while enabling controlled exploration of uncertainty, learning, safety, licensing, and insurance adders without destabilizing the standards-aligned core. This includes the web-based interface for the python code (library/package).

\subsection{Reproducible reporting and traceability via template-driven \LaTeX{} generation}
\label{sec:latex_traceability}

A core design objective of the workflow is that every published cost total is traceable to (i) explicit intermediate computed quantities (e.g., $P_{E,\mathrm{net}}$, component volumes/areas, driver electrical demand), (ii) explicit roll-ups into standardized cost accounts, (iii) each cost account explicitly stating the cost basis (or assumption) for the cost calculation. The report-generation layer implements this by copying \LaTeX{} templates into a build directory and overwriting placeholder strings with computed values. This approach ensures that the power-balance tables, cost-account tables, and LCOE outputs are mechanically linked to the computed state, and that the mapping from subsystem quantities to account totals remains explicit and auditable. 

\subsection{Probabilistic costing: compounding uncertainty from materials, maturity, and learning}
\label{sec:prob_costing}

Point estimates are necessary for comparability, but they are insufficient for decision-making in settings where technology maturity, supply chains, and regulatory pathways are evolving. The development therefore introduces a probabilistic costing layer intended to quantify uncertainty in three dominant sources and then compound them into distributions for subsystem and plant-level costs:
\begin{enumerate}
  \item \textbf{Materials cost uncertainty} (market volatility, purity/specification uncertainty, and supplier depth),
  \item \textbf{Technology maturity uncertainty} represented by Technology Readiness Level (TRL),
  \item \textbf{Learning uncertainty} in experience-curve slopes (both the mean learning rate and its confidence bounds).
\end{enumerate}
The baseline cost-account rollups provide the deterministic ``mean-case'' reference; the probabilistic layer treats each uncertainty source as a multiplicative factor applied to the relevant subset of accounts, enabling explicit attribution of uncertainty to (e.g.) magnets, blankets, power supplies, or safety-critical systems.

\paragraph{Uncertainty propagation.}
Let $C_0$ denote a baseline cost element from the standards-aligned account model (e.g., a subaccount within Category 22). We represent uncertain cost as
\begin{equation}
  C = C_0 \, U_{\mathrm{mat}} \, U_{\mathrm{TRL}} \, U_{\mathrm{LR}},
\end{equation}
where $U_{\mathrm{mat}}$ captures material price/spec uncertainty, $U_{\mathrm{TRL}}$ captures maturity-related dispersion, and $U_{\mathrm{LR}}$ captures learning-rate uncertainty for elements expected to benefit from repetition, modularization, or scaled manufacturing. Each $U_i$ is represented as a probability distribution (typically log-normal for strictly-positive factors), and Monte Carlo sampling yields an empirical distribution for $C$ and for higher-level rollups (e.g., $C_{22}$, $C_{20}$, total capitalized cost).

\subsubsection{Materials cost uncertainty model}
\label{sec:materials_uncertainty}

For material-dominated components (first wall, blanket, shield, structural steels, specialty conductors), the code represents unit material cost as a stochastic variable parameterized either from historical price series (when available) or from expert-specified mean/variance pairs for specialty grades. Component material mass is computed from geometry-driven volume estimates and density:
\begin{equation}
  m = \rho V,
\end{equation}
and material-cost contribution is
\begin{equation}
  C_{\mathrm{mat}} = m \, c_{\mathrm{unit}},
\end{equation}
with $c_{\mathrm{unit}}$ modeled probabilistically (e.g., log-normal) to reflect skew and heavy tails observed in commodity and specialty-material markets. This approach is designed to be consistent with the baseline account model: the stochastic layer does not replace the account taxonomy, but rather provides uncertainty intervals on the same account totals used for deterministic reporting.

\subsubsection{TRL-based maturity uncertainty model}
\label{sec:trl_uncertainty}

Technology maturity is treated as a driver of dispersion around an expected subsystem cost. For a given subsystem with expected cost $C_{\mathrm{exp}}$, we assign a TRL-dependent uncertainty factor with decreasing variance as TRL increases. In the CATF IWG prototype, TRL controls the standard deviation of a strictly-positive distribution for the cost multiplier, e.g.
\begin{equation}
  U_{\mathrm{TRL}} \sim \mathrm{LogNormal}\!\left(\mu(\mathrm{TRL}),\sigma(\mathrm{TRL})\right),
\end{equation}
where $\sigma(\mathrm{TRL})$ decreases monotonically from low TRL (conceptual) to high TRL (commercial). This permits consistent comparisons between concepts at different maturity levels while maintaining a shared deterministic baseline for like-for-like reporting.

\subsubsection{Learning-rate uncertainty estimation via bootstrapped experience-curve regression}
\label{sec:learning_uncertainty}

Learning effects are represented using the standard experience-curve form
\begin{equation}
  C(N) = C_1 N^{b},
\end{equation}
where $N$ is cumulative production (or cumulative installed capacity) and $b<0$ is the learning exponent. The learning rate (fractional cost reduction per doubling) is
\begin{equation}
  \mathrm{LR} = 1 - 2^{b}.
\end{equation}
To quantify uncertainty in $b$, the CATF IWG prototype fits a linear regression to log-transformed data,
\begin{equation}
  \ln C = \ln C_1 + b \ln N,
\end{equation}
and then bootstraps the $(\ln N, \ln C)$ pairs to generate an empirical distribution of $b$ (and therefore LR). This produces confidence intervals for learning assumptions used in cost-out scenarios (e.g., modularized pulsed-power assemblies, repeatable magnet modules, or standardized plant buildings). The resulting distribution enters the propagation model through $U_{\mathrm{LR}}$.

\subsection{Driver-centric extension of the ARPA-E costing capability across MFE, IFE, and MIFE}

A central outcome of the CATF IWG costing cycle was to strengthen \emph{driver-centric} costing so that the dominant fusion-unique subsystem is treated explicitly and traceably for each architecture class. In the standards-aligned code-of-accounts (COA) used by the pyFECONs toolchain, the dominant fusion power core (Cost Account '22. Heat Island') cost driver has historically been represented in Account~22.1.3 (``Coils or Lasers or Pulsed Power''). During the IWG cycle, this generic placeholder was treated as an intentional \emph{swap-point}: for each architecture family, Account~22.1.3 is replaced by a dedicated, bottom-up cost account development for the appropriate driver---\textbf{magnets for MFE}, \textbf{lasers for IFE}, and \textbf{pulsed power for MIFE}. This refactor preserves a common COA spine for balance-of-plant and shared systems, while enabling like-for-like cross-architecture comparisons with auditable traceability from requirements to rolled-up costs.

A key design principle is that the driver account development is not merely a single scaling curve. Instead, it is a structured cost model with (i) requirements-to-quantities logic, (ii) a documented cost basis (e.g., \$/kg, \$/m, \$/W, \$/J, vendor-informed benchmarks), (iii) installation and integration treatment, and (iv) lifetime/replacement logic where relevant.  In each cost account, we model with as much fidelity as is possible, down to the 5th level (e.g. 22.1.3.1.3.5) sometimes further if there is a cost driver and information available on it.  The following three subtopics summarize how the IWG implemented (and intends to continue implementing) this approach.

\subsubsection{Magnetic Fusion Energy (MFE): replacing Account~22.1.3 with an explicit magnet cost account development}

For MFE concepts, magnets are typically the dominant fusion-unique capital driver. The IWG therefore treats Account~22.1.3 as \emph{``Magnets''} in MFE, replacing the generic ``Coils or Lasers or Pulsed Power'' placeholder with a magnet-focused cost account development. The intent is to connect physics and engineering requirements (field strength, bore size, stored energy, coil set geometry, structural fraction, and allowable stresses) to explicit costed quantities (superconductor, conductor, structural materials, cryostat/cryogenics interfaces, and manufacturing steps).

The magnet cost account development is structured to support:
\begin{itemize}
  \item \textbf{Materials and bill-of-materials logic:} explicit superconducting material quantities (e.g., HTS/LTS), stabilizer and structural materials, and associated fabrication yields where available.
  \item \textbf{Manufacturing and integration:} coil winding/stacking, impregnation, joints/terminations, cryostat integration interfaces, and installation labor factors represented as explicit elements rather than embedded in a single multiplier.
  \item \textbf{Design variants and sensitivity:} the framework is designed to let the user vary the dominant cost levers (e.g., HTS cost assumptions, structural mass fractions, winding pack current density limits, coil segmentation strategy) and observe impacts on the rolled-up plant cost.
\end{itemize}

In this architecture, Account~22.1.7 (``Power Supplies'') remains important but secondary: it captures conventional electrical plant equipment supporting magnet operation, while Account~22.1.3 becomes the primary, magnet-specific fusion driver account for MFE.

\subsubsection{Inertial Fusion Energy (IFE): replacing Account~22.1.3 with an explicit laser and target-drive cost account development}

For IFE concepts, the dominant driver is the laser (or more generally the driver delivering energy to the target). The IWG therefore treats Account~22.1.3 as \emph{``Lasers / Driver''} in IFE, replacing the generic placeholder with a laser (or projectile or beam)-focused cost account development that is anchored to delivered energy per shot, repetition rate, wall-plug efficiency, and modular driver architecture.

The IFE driver account development emphasizes:
\begin{itemize}
  \item \textbf{Requirements-to-plant sizing:} linking pulse energy, repetition rate, and efficiency to required input electrical power, thermal management, and driver module count.
  \item \textbf{Driver module costing:} representing the laser system as a set of repeatable modules (e.g., gain media, pumping, optics, beam transport, controls, thermal management), with the ability to map vendor/benchmark costs to each element.
  \item \textbf{Consumables and high-cycle operations:} providing a clear interface between the laser/driver account and high-throughput operational elements (e.g., optics lifetime assumptions, replacement schedules) where these materially affect lifecycle cost.
\end{itemize}

As in MFE, Account~22.1.7 remains the electrical support category, but for IFE the driver itself (laser) is the principal fusion-unique account. The IWG’s benchmarking work on laser targets and driver-adjacent subsystems is intended to tighten this cost basis and reduce reliance on coarse historical scalings.  In all of these cost accounts, we are applying learning curves to the dominant costs to understand the impact of modularity and the scaling of costs to many modules.

\subsubsection{Magneto-Inertial Fusion Energy (MIFE): replacing Account~22.1.3 with an explicit pulsed-power cost account development}

For MIFE concepts, pulsed power is frequently the dominant capital driver and is not well represented by generic electrical-plant scalings. The IWG therefore treats Account~22.1.3 as \emph{``Pulsed Power (Driver)''} in MIFE, replacing the generic placeholder with a bottom-up pulsed-power cost account development. The implementation is anchored to installed \$/J (stored energy) cost bases, explicit module counts, and lifetime/replacement logic driven by shot rate and component derating assumptions.

The MIFE pulsed-power account development includes:
\begin{itemize}
  \item \textbf{Component-level decomposition:} capacitors, switches, charging, dump circuits, buswork/harnessing, diagnostics, safety and controls represented as explicit cost elements (enabling structured sensitivity and vendor refresh).
  \item \textbf{Installed \$/J basis with explicit scaling:} the total pulsed-power cost scales transparently with required stored energy and the number of replicated modules/stages, rather than embedding replication inside opaque scaling factors.
  \item \textbf{Lifetime and derating logic:} explicit switching between (i) a replacement regime (shot-limited component lifetimes) and (ii) a life-of-plant regime enabled by derating/overbuild, converting recurring replacements into higher upfront capex with clearer implications for availability and lifecycle cost.
\end{itemize}

In MIFE, Account~22.1.7 (``Power Supplies'') remains the natural home for much of the pulsed-power infrastructure detail (charging systems, power conditioning, and related plant electrical equipment). However, the \emph{driver-centric} refactor is implemented by making Account~22.1.3 itself the MIFE-specific pulsed-power driver account, with a clear and explicit linkage to 22.1.7 for supporting electrical infrastructure. This preserves COA consistency across architectures while reflecting the reality that pulsed power is a first-order fusion-unique cost driver in MIFE.

\paragraph{Summary.}
Across all three architecture families, the IWG’s future implementation plan is to treat Account~22.1.3 as a controlled ``driver swap'' that is replaced by a full, auditable cost account development: magnets (MFE), lasers (IFE), and pulsed power (MIFE). This provides a consistent COA backbone for balance-of-plant while enabling the fidelity needed to evaluate fusion-unique cost drivers on equal methodological footing.

\subsection{Extension of the ARPA-E costing capability for learning (experience curves) and its application to magnets and lasers}

A second major extension implemented in the CATF-supported costing framework is explicit treatment of learning (experience curves), both as a deterministic cost-out lever for NOAK-style projections and as a quantified uncertainty contributor in probabilistic costing. The goal is to move learning from qualitative narrative to a transparent, parameterized component of the costing workflow, with clear mapping to the cost accounts expected to benefit from repetition, modularization, and scaled manufacturing.

\subsubsection{Learning-curve foundations and parameterization}

The learning module adopts Wright’s curve as a canonical representation of cost reduction with cumulative production:
\begin{equation}
Y = a X^{b},
\end{equation}
where $Y$ is average unit cost, $X$ is cumulative units produced, $a$ is first-unit cost, and $b<0$ is the learning elasticity \cite{WoodruffLearning}. The learning rate (fractional cost reduction per cumulative doubling) is represented in the framework using the standard relationship
\begin{equation}
LR = 1 - 2^{b},
\end{equation}
which provides a direct mapping between an interpretable ``percent improvement per doubling'' parameter and the exponent used in cost calculations \cite{LearningUncBootstrapping}. This formulation supports consistent application across fusion subsystems that may plausibly experience manufacturing learning (e.g., modular pulsed-power assemblies, repeatable magnet modules, standardized buildings) \cite{LearningUncBootstrapping}.

\subsubsection{Implementation as an uncertainty factor compounded with materials and maturity}

In the CATF IWG probabilistic costing layer, learning is treated as one of three dominant multiplicative uncertainty sources that compound into distributions for subsystem and plant-level costs:
\begin{equation}
C = C_{0}\,U_{\mathrm{mat}}\,U_{\mathrm{TRL}}\,U_{\mathrm{LR}},
\end{equation}
where $C_{0}$ is the deterministic baseline cost element, and $U_{\mathrm{mat}}$, $U_{\mathrm{TRL}}$, and $U_{\mathrm{LR}}$ are stochastic multipliers representing materials-cost uncertainty, technology maturity (TRL) uncertainty, and learning-rate uncertainty, respectively \cite{ProbCostingEq}. The learning-rate uncertainty is estimated by fitting a log-linear regression to experience-curve data and bootstrapping $(\ln N,\ln C)$ pairs to generate an empirical distribution of the learning exponent $b$ (and therefore of $LR$). The resulting distribution enters the Monte Carlo cost propagation through $U_{\mathrm{LR}}$ \cite{LearningUncBootstrapping}. This construction is deliberately aligned with the COA: uncertainty is not ``added on'' outside the accounting structure, but applied to selected accounts (e.g., magnets, power supplies) so that uncertainty attribution remains explicit and auditable \cite{ProbCostingEq}.

\subsubsection{Application of learning to magnets (MFE), lasers (IFE), and pulsed power (MIFE)}

A central motivation for adding explicit learning is the recognition that cost-out trajectories for fusion are likely to be
dominated by the learning behavior of the \emph{primary driver technologies}. In the ARPA-E framework, these driver
technologies are handled in a deliberately consistent way across architectures by using a common COA ``swap-point''
for the dominant fusion-unique driver while preserving the same reporting envelope for balance-of-plant and shared
accounts. In particular, the concept-defining driver for MFE, IFE, and MIFE is mapped to Account~22.1.3
(``Coils or Lasers or Pulsed Power''), with supporting electrical infrastructure captured in Account~22.1.7
(``Power Supplies'') and installation/assembly represented explicitly in the appropriate installation subaccounts
(e.g., 22.01.11.03) \cite{COA,DriversAcrossArch}. This architecture enables learning and uncertainty to be applied
directly to the accounts that dominate total cost and drive the widest uncertainty.

\begin{itemize}
  \item \textbf{Magnets for MFE (and some MIF variants):} primary magnet/coil costs (TF, PF, and CS for tokamaks) are
  mapped to Account~22.1.3, with installation/assembly treated explicitly (e.g., 22.01.11.03 ``Magnets'') \cite{COA}.
  Learning can be applied to repeatable magnet modules and manufacturing steps where volume production is plausible
  (e.g., conductor procurement, winding/stacking, impregnation, joints/terminations, cryostat interfaces, and factory
  acceptance testing). Uncertainty can be attributed directly to magnet-related accounts so that the effect of magnet
  learning assumptions remains traceable in the rolled-up COA.

  \item \textbf{Lasers for IFE:} the driver is a dominant, concept-defining element requiring explicit modeling and
  appears in the same Account~22.1.3 location for COA comparability across MFE/IFE/MIFE \cite{DriversAcrossArch}.
  Learning assumptions can therefore be applied consistently to laser/driver module manufacturing and assembly where
  repetition and scaled production are credible (e.g., gain modules, pumping, optics, beam transport, controls, and
  thermal management), with uncertainty similarly mapped to the driver accounts.

  \item \textbf{Pulsed power for MIFE:} the dominant driver is the pulsed-power system (often implemented as capacitor
  banks, switching, charging, dump circuits, buswork/harnessing, controls, and protection). In the same COA structure,
  the MIFE driver is treated as the Account~22.1.3 swap-in (``Pulsed Power''), while major supporting electrical plant and
  pulse-forming infrastructure remain naturally represented in Account~22.1.7 (``Power Supplies'') \cite{COA,DriversAcrossArch}.
  This division is important because pulsed power is both \emph{highly modular} and \emph{highly repetition-sensitive}:
  credible learning can be applied to replicated modules (banks, switch assemblies, charger units, and standardized
  interconnects), and learning/uncertainty can be separated between (i) driver hardware (22.1.3) and (ii) supporting
  power-conditioning infrastructure (22.1.7) when their supply chains and learning rates differ. This enables MIFE
  cost-out scenarios to reflect realistic design-for-manufacture trajectories (standardization, increased production
  volume, improved yields, and reduced installation labor) without obscuring the accounting traceability.

\end{itemize}

This design ensures that learning-driven cost-out scenarios for magnets and lasers are represented in the same
standards-aligned reporting envelope as learning-driven cost-out scenarios for MIFE pulsed power. It also preserves a
key methodological principle emphasized in the ARPA-

\subsection{Safety-informed costing: hazard mapping to cost-account adders and safety-driven plant provisions}
\label{sec:safety_costing}

This paper extends the baseline standards-aligned costing workflow by incorporating a \emph{safety-informed scoping analysis}. The safety module is designed to (i) make fusion-specific hazards explicit at the pre-conceptual and conceptual design stages, (ii) translate those hazards into first-order \emph{amelioration measures} (systems, structures, components, and operational provisions), and (iii) ensure that the resulting scope is \emph{mapped into the standardized cost accounts} rather than treated as an untracked adder.  Not all the hazards have been translated into costs, and the work is ongoing at the time of writing.  We draw heavily from the work of White 
\cite{White2021FusionLicensing} and Lukacs and Williams \cite{LukacsWilliams2021HazardSensitivity}.

\subsubsection{Purpose and scope of the safety module}
\label{sec:safety_scope}

The safety analysis implemented is not a substitute for full probabilistic risk assessment, licensing-grade safety cases, or detailed shielding/transport calculations. Instead, it is a \emph{costing-oriented} module that:
\begin{enumerate}
  \item enumerates representative hazards relevant to commercial fusion facilities,
  \item provides simplified screening metrics and sizing calculations for mitigating features, and
  \item produces cost-relevant quantities (mass, area, thickness, footprint, capacity, or proxy costs) that can be allocated to cost accounts.
\end{enumerate}
This structure supports consistent early-phase comparisons between concepts while preventing safety scope from being omitted or double-counted.

\subsubsection{Hazard taxonomy}
\label{sec:hazard_taxonomy}

The current implementation includes hazard descriptions (and, where available, scoping calculations) for the following classes:
\begin{itemize}
  \item \textbf{Plasma disruption and off-normal plasma behavior} (thermal loads, induced currents, and magnetic forces on first wall / blanket structures).
  \item \textbf{Rapid structural oxidation} following vacuum boundary breaches and oxidant ingress, including first-wall protection measures (e.g., coatings).
  \item \textbf{Direct radiation exposure} and \textbf{bioshielding} needs, including a dose-transmission-factor (DTF) approach to estimating required shield thickness.
  \item \textbf{Cooling disruption accidents} (LOCA/LOFA/LOHR) and afterheat/continued heating considerations, including screening-type ``safety margin'' calculations against damage/melt thresholds.
  \item \textbf{Corrosion/mass transport} and \textbf{radioactivity mobilization} in coolants and breeder/blanket systems, represented through relative consequence indices.
  \item \textbf{Cryogenic hazards} (asphyxiants, large coolant inventories) affecting plant access controls, detection, and ventilation provisions.
  \item \textbf{Tritium and activated material releases} (e.g., tritium in the form of HTO and dust mobilization), including scoping of site-boundary implications and the systems needed for controlled releases (suppression/vent and detritiation).
  \item \textbf{Supplementary heating hazards} (high-energy neutral/ion beam exposure risks and interlocks).
  \item \textbf{Vacuum system hazards} (activated gases, tritium-bearing streams, erosion products).
  \item \textbf{Radwaste hazards and handling} (classification, storage, recycling/detritiation, and long-term disposition).
  \item \textbf{Fuel handling hazards} (tritium systems and associated contamination control and confinement).
\end{itemize}

\subsubsection{Screening metrics and sizing calculations implemented}
\label{sec:safety_calcs}

The notebook includes several calculation patterns that translate qualitative hazards into scoping-level design and cost drivers:

\paragraph{Relative consequence indices (RCIs).}
For selected hazards, the implementation uses normalized indices to screen the severity of a condition relative to a reference/best case. Examples include:
\begin{itemize}
  \item an oxidation-related RCI based on separation between operating temperature and onset/limit temperatures for candidate structural materials;
  \item corrosion/mass-transport and release-related RCIs based on representative corrosion rates and releasable inventory proxies.
\end{itemize}
These indices are used to motivate whether an amelioration feature should be included (and to support comparative sensitivity across material/coolant choices).

\paragraph{First-wall protection cost proxy (rapid oxidation amelioration).}
A worked example estimates tungsten coating mass from first-wall surface area and coating thickness, then builds a simple total cost including material cost plus application and R\&D multipliers. This provides an explicit cost hook for oxidation protection measures rather than embedding them implicitly in blanket/first-wall base costs.

\paragraph{Bioshield thickness and cost proxy (direct radiation exposure).}
The bioshield module uses dose transmission factors (DTFs) versus thickness for candidate shielding materials, fits a simple relationship, and then determines the thickness required to reduce an initial annual dose to a regulatory dose target via:
\begin{equation}
  \mathrm{Dose_{out}} = \mathrm{Dose_{in}} \times \mathrm{DTF},
  \qquad
  \mathrm{DTF_{req}} = \frac{\mathrm{Dose_{limit}}}{\mathrm{Dose_{in}}}.
\end{equation}
A material-density and $/kg$ approach then provides a first-order cost for the shield mass/volume implied by the required thickness.

\paragraph{Cooling-disruption screening (LOCA/LOFA/LOHR).}
For disruption scenarios, the notebook includes a simplified ``safety margin'' construct comparing deposited heat over shutdown times to material heat capacity and allowable temperature rise thresholds. While not licensing-grade, this creates an auditable link between shutdown time assumptions, first-wall thickness/material, and the need for additional protective design or operational controls.

\paragraph{Site-boundary implications of tritium/dust release.}
A scoping model uses inventory-versus-distance relationships (parameterized from literature data embedded in the notebook) to estimate site boundary distances that would trigger sheltering or evacuation for given mobilizable inventories. The code then computes implied site area and a proxy site cost based on acreage.

\paragraph{Radwaste and detritiation proxies.}
The implementation includes (i) a detritiation system proxy cost (e.g., vacuum oxygen decarburization as a representative detritiation/recycling step) and (ii) a radwaste storage cost proxy based on an annual \$/kg-year storage rate and assumed waste mass and storage duration.

\subsubsection{Mapping safety scope into standardized cost accounts}
\label{sec:safety_mapping}

The key methodological contribution is the \emph{mapping} of each hazard’s mitigating features into the standardized cost accounts used throughout the costing framework. In practice, safety measures manifest in (a) additional or resized plant structures/buildings, (b) additional or resized fusion-island and fuel-cycle equipment, (c) site/exclusion-zone and boundary provisions, and (d) regulatory/financial adders that depend on the assumed hazard posture.

Below is the account mapping used for integrating the safety module outputs. Account names follow the same chart-of-accounts convention used elsewhere in this paper.

\begin{itemize}
  \item \textbf{Account 20: Capitalized direct costs (site/land implications).}
  \begin{itemize}
    \item Site boundary / exclusion zone implications (when treated as a land or site acquisition driver).
  \end{itemize}

  \item \textbf{Account 21: Buildings (structures and site facilities).}
  \begin{itemize}
    \item Machine building and confinement-related structures.
    \item Bioshield / shielding structures whose dominant expression is as building concrete/structures.
    \item Vent/stack routing and building-integrated confinement volumes (as applicable).
  \end{itemize}

  \item \textbf{Account 22.1: Fusion island equipment (bottom-up components).}
  \begin{itemize}
    \item First-wall / blanket / shield modifications driven by safety (e.g., protective coatings, structural provisions).
    \item Vacuum vessel and penetrations with confinement and accident-mitigation features.
    \item Primary confinement boundary components affected by oxidation/cooling-disruption posture.
    \item Shielding treated as part of the fusion island equipment rather than building structure (when appropriate).
  \end{itemize}

  \item \textbf{Account 22 (other Heat Island plant equipment, including fuel cycle and radioactive systems).}
  \begin{itemize}
    \item Tritium handling, suppression/vent systems, and detritiation systems.
    \item Vacuum pumping streams handling activated gases and tritium-bearing effluents.
    \item Radwaste handling, interim storage, recycling/detritiation steps, and contamination control equipment.
    \item Cryogenic systems where treated as plant equipment (detection, ventilation interfaces, inventory minimization provisions).
    \item Supplementary heating safety interlocks and access controls (where treated as equipment scope).
  \end{itemize}

  \item \textbf{Account 52: Contingencies (supplementary costs).}
  \begin{itemize}
    \item Safety-driven uncertainty reserves when hazard posture is not yet resolved at conceptual stage (optional treatment; may be applied selectively to safety-sensitive scope).
  \end{itemize}

  \item \textbf{Account 53: Insurance (supplementary costs).}
  \begin{itemize}
    \item Insurance premiums or capitalized insurance provisions, scenario-parameterized by assumed risk posture and hazard controls.
  \end{itemize}

  \item \textbf{Account 54: Decommissioning (supplementary costs).}
  \begin{itemize}
    \item Safety-relevant end-of-life liabilities (radwaste disposition, activated component handling) when treated as decommissioning scope.
  \end{itemize}

  \item \textbf{Account 62: Fees (financial costs).}
  \begin{itemize}
    \item Licensing and regulatory fees that depend on the safety posture, licensing pathway, and jurisdictional requirements.
  \end{itemize}
\end{itemize}

\subsubsection{Hazard-to-account mapping summary}
\label{sec:hazard_to_account_table}

Table~\ref{tab:hazard_to_account} summarizes how each hazard class in the CATF IWG notebook maps to cost-account locations. This mapping is intended to be conservative and auditable: each safety-driven feature is recorded where it would be procured/constructed in an EPC sense (buildings vs.\ equipment vs.\ supplementary/financial adders).

\begin{table}[h]
\centering
\caption{Safety module hazard classes, representative ameliorations, and cost-account mapping.}
\label{tab:hazard_to_account}
\begin{tabular}{p{0.28\linewidth} p{0.44\linewidth} p{0.22\linewidth}}
\hline
\textbf{Hazard class} & \textbf{Representative ameliorations in the module} & \textbf{Primary account mapping} \\
\hline
Plasma disruption / off-normal plasma behavior &
Structural margin provisions; robustness of first wall/blanket; operational/shutdown constraints &
22.1 (fusion island) \\
Rapid structural oxidation &
First-wall protection (e.g., coating proxy); confinement/penetration robustness &
22.1; (21 if building confinement changes) \\
Direct radiation exposure &
Bioshield thickness and cost proxy (material-dependent) &
21 (bioshield as structure) and/or 22.1 (shield as equipment) \\
Cooling accidents (LOCA/LOFA/LOHR) &
Safety-margin screening; additional protection/cooling provisions where implied &
22.1; (22 if dedicated safety systems are added) \\
Corrosion / activated transport &
Screening via RCIs; filtration/cleanup implications; contamination control emphasis &
22 (radioactive systems); 54 (end-of-life implications) \\
Cryogenic hazards &
Inventory minimization; detection/ventilation interfaces; access controls &
22 (plant equipment); 21 (building interfaces) \\
Tritium/dust release and site boundary &
Suppression/vent + detritiation; site boundary sizing and proxy site cost &
22 (detritiation/suppression); 20/21 (site boundary/structures) \\
Supplementary heating beam hazards &
Interlocks, shielding/access control provisions on beamlines &
22 (plant equipment); 21 (shielding/access areas) \\
Vacuum system radioactive streams &
Contamination control, filtration, tritium-compatible effluent handling &
22 (radioactive systems) \\
Radwaste handling and storage &
Detritiation/recycling proxy; interim storage proxy and duration assumptions &
22 (radwaste systems); 54 (decommissioning linkage) \\
Fuel handling hazards &
Tritium systems, confinement and contamination control, processing equipment &
22 (fuel cycle / radioactive systems) \\
Licensing and compliance implications &
Licensing fees and jurisdictional compliance processes (scenario-driven) &
62 (fees) \\
Insurance implications &
Risk-posture-dependent insurance premiums/provisions &
53 (insurance) \\
\hline
\end{tabular}
\end{table}

\subsubsection{Implementation note}
\label{sec:safety_impl_note}

In the current codebase, safety calculations are implemented as modular, worked-example computations (e.g., coating mass/cost; bioshield thickness from DTF targets; site boundary from inventory-distance relationships; radwaste and detritiation proxies). Usually the work flow in costing analysis is to consider the specific embodiment for each configuration individually, replacing stand-alone proxy values with (i) plant-specific geometry and inventory outputs from the core plant model and (ii) explicit allocation of computed costs into the corresponding cost accounts (Accounts 20--22 for direct scope; Accounts 52--54 for supplementary costs; and Account 62 for licensing/fees).  Currently the cost calculations are included in the appendices and are referred to as the cost bases for the cost accounts that use them, keeping the standard as clean as possible.
A specific CATF-supported evolution of the framework is expanded treatment of safety systems and safety-driven plant provisions within the standardized account taxonomy \cite{CATFcolab}. In the CATF IWG prototype notebook, the safety methodology is structured around two coupled elements:

\paragraph{(1) Hazard identification and mapping.}
Representative internal hazards are enumerated (e.g., plasma disruptions, coolant accidents, cryogenic hazards, tritium release scenarios, supplementary heating hazards, vacuum/radwaste/fuel-handling hazards, and hot-cell/remote-handling radioactivity). Each hazard is mapped to the systems, structures, and components (SSCs) that provide prevention, mitigation, and consequence limitation (e.g., confinement boundaries, detritiation, shielding/bioshield, drainage/suppression systems, remote handling, radwaste systems). These SSCs, in turn, map naturally into the standardized cost accounts (primarily within Category 22 and supporting structures/buildings in Category 21), preserving comparability while making fusion-specific safety scope explicit.

\paragraph{(2) Safety-driven sizing rules and capital provisions.}
Where hazards imply geometric or shielding requirements, we compute first-order sizing provisions within the same geometry-driven plant model used for the heat island (e.g., bioshield thickness and associated volumes/areas). These sizing provisions then feed into the same account rollups used in the baseline model. The intent is not to claim final safety design, but to ensure that safety-driven scope is not omitted from early cost estimates, and that safety adders appear in auditable, traceable account locations.

\subsection{Regulatory and financial adders: licensing fees and insurance premiums}
\label{sec:regulatory_insurance}

The standardized account structure explicitly includes licensing fees (Category 62) and insurance (Category 54) \cite{CATFcolab}. CATF IWG development prototyped methods to estimate these adders in a way that can be scenario-parameterized by jurisdiction and by assumed safety posture, also referred to as the cost basis and full discussion included as appendices to the standard report.

\subsubsection{Licensing cost proxy model}
\label{sec:licensing_model}

The prototype licensing model represents total licensing cost as the sum of (i) labor-hour-driven regulatory review fees (hours by application type $\times$ hourly fee schedule) and (ii) fixed fees associated with byproduct materials or other regulatory categories, conditional on country/jurisdiction. In simplest form,
\begin{equation}
  C_{\mathrm{lic}} = r_{\mathrm{hour}} \sum_{k} H_k + C_{\mathrm{fixed}},
\end{equation}
where $H_k$ are the labor hours associated with licensing actions (e.g., early site permits, combined licenses, design certifications, amendments) and $r_{\mathrm{hour}}$ is the regulatory hourly rate. This structure supports sensitivity studies in which fusion-specific licensing pathways differ from fission, while still allocating the result into the ... standardized Category 62 account for consistent reporting \cite{White2021FusionLicensing}.

\subsubsection{Insurance premium scaling model}
\label{sec:insurance_model}

Insurance is treated as an annual recurring obligation with a capitalized or annualized representation depending on reporting preference. The CATF IWG prototype uses a comparative-risk approach: premiums are scaled relative to a reference (e.g., a fission benchmark premium) using a risk or safety-rating proxy (e.g., a safety assurance level or risk index derived from hazard posture and consequence controls). In generic form,
\begin{equation}
  P_{\mathrm{fusion}} = P_{\mathrm{ref}} \, \phi(\mathcal{R}),
\end{equation}
where $P_{\mathrm{ref}}$ is a reference premium and $\phi(\mathcal{R})$ is a scaling factor derived from a risk proxy $\mathcal{R}$. This provides an auditable pathway to include insurance in Category 54 and to test how improved hazard controls (better confinement, reduced inventory at risk, stronger mitigation systems) translate into reduced insurance burden.

\subsection{Beyond LCOE: investment and value metrics, and time-dependent market contexts}
\label{sec:value_metrics}

This section documents the \emph{economics measures module} developed in the CATF IWG, intended to complement the baseline standards-aligned cost-account and LCOE workflow by enabling value-based metrics computed from the same cost and performance outputs.  These calculations are included as appendices and only exercised when the full standard costing is completed. The measures enumerated and/or implemented in the CATF IWG code include: Net Present Value (NPV), Total Life-Cycle Cost (TLCC), Revenue Requirements, Levelized Cost of Energy (LCOE), Annualized Value (equivalent annual cost/value), Internal Rate of Return (IRR), Modified Internal Rate of Return (MIRR), Simple and Discounted Payback Period, Benefit-to-Cost Ratios, Savings-to-Investment Ratios, Integrated Resource Planning / Demand-Side-Management (IRP/DSM) ratio tests, Consumer/Producer Surplus, and Weighted Average Cost of Capital (WACC). The module also includes asset tail-value metrics (Residual Value and Follow-On Value) and an alternative valuation framework (Adjusted Present Value, APV).  Note that many of these metrics have been captured in a web-interface \cite{Weeks2026ForecastingFusionMarketplace}.

\subsubsection{Common inputs and notation}
\label{sec:econ_common}

We define:
\begin{itemize}
  \item $I_0$: initial capital outlay at $t=0$ (e.g., total capital cost or overnight cost adjusted to the chosen base year).
  \item $t \in \{1,\dots,T\}$: year index over the financial evaluation horizon $T$.
  \item $CF_t$: net cash flow in year $t$ (after-tax if desired), typically
  \begin{equation}
    CF_t = R_t - C_t,
  \end{equation}
  where $R_t$ is revenue and $C_t$ includes O\&M, fuel, replacement/refurbishment, insurance, licensing, and other annual costs.
  \item $r$: discount rate (nominal or real, consistently applied). In this module $r$ is commonly taken as WACC (Section~\ref{sec:wacc}).
  \item $E_t$: net electrical energy delivered in year $t$ (MWh/y), derived from net power and availability.
\end{itemize}
Unless otherwise stated, present values use the standard discount factor $(1+r)^{-t}$.

\subsubsection{Levelized cost of energy (LCOE)}
\label{sec:lcoe}

LCOE is retained as the baseline comparability metric and is computed from annualized cost divided by annual delivered energy. Let $\mathcal{A}$ denote the total annualized cost (fixed charge on capital plus annual costs):
\begin{equation}
  \mathrm{LCOE} = \frac{\mathcal{A}}{E},
\end{equation}
where $E$ is annual net energy delivered (MWh/y) and $\mathcal{A}$ aggregates annualized capital and recurring costs. When a constant annualization convention is used, the annualized capital component is commonly expressed via the capital recovery factor (CRF):
\begin{equation}
  \mathrm{CRF}(r,T) = \frac{r(1+r)^T}{(1+r)^T - 1},
  \qquad
  \mathcal{A}_\mathrm{cap} = I_0 \, \mathrm{CRF}(r,T).
\end{equation}

\subsubsection{Net present value (NPV)}
\label{sec:npv}

The code implements NPV as the discounted sum of future cash flows relative to the initial investment:
\begin{equation}
  \mathrm{NPV}(r) = -I_0 + \sum_{t=1}^{T} \frac{CF_t}{(1+r)^t}.
\end{equation}
In implementation, $I_0$ is taken from the plant capital cost totals already computed by the costing framework, while $CF_t$ can be constructed from scenario-specific revenue assumptions (e.g., energy price times $E_t$) minus annual cost streams (O\&M, fuel, replacements, insurance, licensing, etc.). This enables direct comparison of fusion concepts on a value basis while preserving traceability to the underlying cost accounts and operating assumptions.

\subsubsection{Internal rate of return (IRR)}
\label{sec:irr}

IRR is computed as the discount rate $r^\ast$ that sets NPV to zero:
\begin{equation}
  0 = -I_0 + \sum_{t=1}^{T} \frac{CF_t}{(1+r^\ast)^t}.
\end{equation}
The code implements IRR by defining the NPV equation and numerically solving for the root (e.g., via a one-dimensional root finder). IRR is reported as an annualized return and is useful for comparing projects with different durations and scale, provided cash-flow sign conventions are consistent.

\subsubsection{Modified internal rate of return (MIRR)}
\label{sec:mirr}

MIRR addresses limitations of IRR by separating (i) the finance rate used to discount negative cash flows and (ii) the reinvestment rate used to compound positive cash flows. Let $r_f$ be the finance rate and $r_r$ the reinvestment rate. Define:
\begin{align}
  \mathrm{PV}_{-} &= \sum_{t=0}^{T} \frac{\min(CF_t,0)}{(1+r_f)^t},\\
  \mathrm{FV}_{+} &= \sum_{t=0}^{T} \max(CF_t,0)\,(1+r_r)^{T-t}.
\end{align}
Then MIRR is:
\begin{equation}
  \mathrm{MIRR} = \left(\frac{\mathrm{FV}_{+}}{|\mathrm{PV}_{-}|}\right)^{1/T} - 1.
\end{equation}
The code follows this structure by partitioning cash flows into positive and negative sets, discounting the negative flows at $r_f$, compounding the positive flows at $r_r$, and computing the implied constant annualized return.

\subsubsection{Weighted average cost of capital (WACC)}
\label{sec:wacc}

The module includes WACC as the default discount rate for NPV-style calculations. With equity $E$, debt $D$, total capitalization $V=E+D$, cost of equity $R_e$, cost of debt $R_d$, and corporate tax rate $\tau$, WACC is:
\begin{equation}
  \mathrm{WACC} = \frac{E}{V}R_e + \frac{D}{V}R_d (1-\tau).
\end{equation}
This provides a consistent bridge between financing assumptions and valuation outputs (NPV, revenue requirements, and annualized value).

\subsubsection{Total life-cycle cost (TLCC)}
\label{sec:tlcc}

TLCC is used to evaluate the total present-value cost of owning and operating the asset over its life. The notebook defines TLCC conceptually as a roll-up of capitalized cost categories plus lifetime recurring costs, less residual value. In general form:
\begin{equation}
  \mathrm{TLCC} = I_0 + \sum_{t=1}^{T}\frac{C_t}{(1+r)^t} - \frac{RV_T}{(1+r)^T},
\end{equation}
where $C_t$ includes O\&M, fuel, replacements, and other annual costs, and $RV_T$ is an end-of-horizon residual/salvage value (if applied at $T$).
The code notes a practical mapping to the costing framework’s account structure by summing:
\begin{itemize}
  \item capital cost totals (direct plus capitalized indirect/owner/supplementary/financial categories),
  \item lifetime operating costs (annual O\&M and annual fuel, multiplied by life and/or discounted year-by-year),
  \item financing costs (either capitalized or annualized consistently),
  \item minus residual value when assets operate beyond a contracted period (Section~\ref{sec:rv}).
\end{itemize}
For comparisons between mutually exclusive alternatives providing the same service, TLCC can be used directly in place of LCOE.

\subsubsection{Revenue requirements}
\label{sec:revenue_req}

Revenue requirements compute the constant annual revenue (or required energy price) needed to meet a financial criterion. Two equivalent formulations are used in this module:

\paragraph{(1) NPV-based revenue requirement.}
Assume an unknown constant annual revenue $R$ (or price $p$ with $R=pE$) and known annual costs $C_t$. Solve for $R$ such that $\mathrm{NPV}(r)=0$:
\begin{equation}
  0 = -I_0 + \sum_{t=1}^{T}\frac{(R - C_t)}{(1+r)^t}.
\end{equation}
When $C_t$ is constant ($C$) and revenue is constant ($R$), this reduces to:
\begin{equation}
  R = C + I_0 \,\mathrm{CRF}(r,T).
\end{equation}

\paragraph{(2) Equivalent annual cost (EAC) formulation.}
Compute the equivalent annual cost (Section~\ref{sec:annualized}) and set revenue equal to EAC (or solve for price using $p = \mathrm{EAC}/E$).

\subsubsection{Annualized value / equivalent annual cost (EAC)}
\label{sec:annualized}

The module includes an annualization step that converts a present-value quantity into a uniform annual stream. Given a present-value cost $\mathrm{PV}$ over horizon $T$, the equivalent annual cost is:
\begin{equation}
  \mathrm{EAC} = \mathrm{PV}\times \mathrm{CRF}(r,T).
\end{equation}
This can be applied to TLCC (to produce an annual cost) or to a project value (to produce an equivalent annual value), enabling consistent comparison with annual revenues, annual costs, and annual energy delivery.

\subsubsection{Payback period}
\label{sec:payback}

The notebook enumerates both simple and discounted payback periods:

\paragraph{Simple payback.}
The smallest $t$ such that cumulative (undiscounted) cash flow becomes non-negative:
\begin{equation}
  \sum_{k=1}^{t} CF_k \ge I_0.
\end{equation}

\paragraph{Discounted payback.}
The smallest $t$ such that cumulative discounted cash flow becomes non-negative:
\begin{equation}
  \sum_{k=1}^{t} \frac{CF_k}{(1+r)^k} \ge I_0.
\end{equation}
These measures are useful as liquidity and risk-screening metrics but do not capture full-life value beyond the payback time.

\subsubsection{Benefit-to-cost ratio (BCR)}
\label{sec:bcr}

Benefit-to-cost ratios compare discounted benefits to discounted costs:
\begin{equation}
  \mathrm{BCR} = \frac{\sum_{t=0}^{T}\dfrac{B_t}{(1+r)^t}}{\sum_{t=0}^{T}\dfrac{K_t}{(1+r)^t}},
\end{equation}
where $B_t$ are benefit streams (e.g., avoided costs, revenues, policy value) and $K_t$ are cost streams (capex and O\&M components, including replacements). BCR is used for screening and ranking when benefits can be monetized consistently across options.

\subsubsection{Savings-to-investment ratio (SIR)}
\label{sec:sir}

Savings-to-investment ratio is treated as a special case of BCR for efficiency-style comparisons, using incremental quantities relative to a reference case:
\begin{equation}
  \mathrm{SIR} = \frac{\sum_{t=0}^{T}\dfrac{\Delta S_t}{(1+r)^t}}{\Delta I_0},
\end{equation}
where $\Delta S_t$ are discounted annual savings (e.g., avoided operating costs) and $\Delta I_0$ is the incremental upfront investment relative to the baseline.

\subsubsection{IRP/DSM ratio tests}
\label{sec:irp_dsm}

The module enumerates IRP/DSM ratio tests as decision metrics used in utility planning contexts. In general, these tests compare discounted avoided costs (benefits) to discounted program and participant costs (costs) using the same present-value structure as BCR, but with different definitions of which cash flows belong to the numerator/denominator depending on perspective (utility system, participant, ratepayer, or societal). In this framework, the relevant inputs (capex, annual O\&M, fuel, replacement/refurbishment, and energy delivery) are already computed by the costing model, enabling consistent construction of the required benefit and cost streams for the chosen planning test.

\subsubsection{Consumer and producer surplus}
\label{sec:surplus}

The module enumerates consumer/producer surplus as market-value measures. In general form, for quantity $Q$ and price $P$ with demand curve $P_D(Q)$ and supply curve $P_S(Q)$:
\begin{align}
  \mathrm{CS} &= \int_{0}^{Q^\ast}\!\!\left(P_D(Q)-P^\ast\right)\,dQ,\\
  \mathrm{PS} &= \int_{0}^{Q^\ast}\!\!\left(P^\ast-P_S(Q)\right)\,dQ,
\end{align}
where $(Q^\ast,P^\ast)$ is the market equilibrium (or the realized operating point under a given dispatch/market assumption). In practical applications within this framework, surplus metrics can be approximated using price-duration and output-duration representations derived from market scenarios, paired with plant net output and marginal operating costs computed by the costing model.

\subsubsection{Residual value (RV)}
\label{sec:rv}

The notebook implements a residual value calculation to capture post-contract economic value when the asset continues operating beyond an initial contracted period. Let $T_c$ be the contracted life and $T_f$ the financial life, and let $NCF$ be net annual cash flow beyond contract:
\begin{equation}
  NCF = R - \mathrm{O\&M}.
\end{equation}
Residual value is computed as the discounted sum of net cash flows for years $t=T_c+1$ to $T_f$ (discounted relative to the contract boundary in the implemented function):
\begin{equation}
  \mathrm{RV} = \sum_{t=T_c+1}^{T_f}\frac{NCF}{(1+r)^{(t-T_c)}}.
\end{equation}
This provides a transparent, scenario-driven representation of tail value consistent with long-lived generation assets.

\subsubsection{Follow-on value (FOV)}
\label{sec:fov}

Follow-on value extends the RV concept to an additional operating period beyond a baseline assumption. In the notebook implementation, FOV is computed as the discounted sum of net annual cash flows over the extension period:
\begin{equation}
  \mathrm{FOV} = \sum_{t=1}^{T_\mathrm{ext}}\frac{NCF}{(1+r)^{t}},
\end{equation}
where $T_\mathrm{ext} = T_\mathrm{actual}-T_c$ is the number of years of operation beyond the contracted life.

\subsubsection{Adjusted present value (APV)}
\label{sec:apv}

The notebook includes APV as an alternative to single-rate NPV when separating project value from financing effects. In standard form:
\begin{equation}
  \mathrm{APV} = \mathrm{PV}\left(\text{unlevered project}\right) + \mathrm{PV}\left(\text{financing side effects}\right),
\end{equation}
where the unlevered project is discounted at the unlevered cost of equity, and financing side effects commonly include the tax shield on debt (discounted at a debt-appropriate rate) and, if modeled, expected costs of financial distress/default. In the notebook implementation, APV includes an unlevered PV term plus a PV tax-shield term constructed from debt expenses and tax rate:
\begin{equation}
  \mathrm{APV} \approx \sum_{t=1}^{T}\frac{CF^{U}_t}{(1+R_e)^t} \;+\; \sum_{t=1}^{T}\frac{\tau \, \mathrm{DebtExp}_t}{(1+R_d)^t}.
\end{equation}

\subsection{Implementation note: integration with the costing pipeline}
\label{sec:econ_integration}

All measures above are designed to be computed using outputs already available in the costing workflow: total capital cost $I_0$ (from the COA rollup), annual O\&M and fuel cost streams (from annualized accounts), replacement/refurbishment schedules (when enabled), net annual energy $E_t$ (from power balance and availability), and scenario-specific revenue assumptions (e.g., fixed price, price-duration curve, or contracted offtake). This ensures that value metrics remain traceable to the same standardized cost accounts used for LCOE reporting, while enabling CATF-relevant sensitivity analyses over discount rates, financing structure, market assumptions, and asset lifetime extensions.



\subsection{Deployment Modeling}
Deployment modeling complements plant-level cost estimates by bounding how quickly fusion can plausibly scale from first-of-a-kind
deployments to meaningful grid penetration. The attached deployment note highlights an \emph{agent-based power-plant fleet model}
that simulates capacity additions as a function of (i) the \emph{fusion market entry date} and (ii) an assumed \emph{market capture ceiling}
for new builds, producing trajectories for installed capacity by technology type (Figure~1 in the note) \cite{Spangher2019ABM}. The key
interpretation for cost analysis is that the commercial opportunity for fusion depends strongly on timing relative to retirements and
load growth: later entry can coincide with larger replacement and growth needs, while a tighter capture ceiling constrains the annual
rate at which fusion can replace incumbent capacity, independent of plant cost. These dynamics motivate coupling pyFECONs-style
costing outputs (FOAK/NOAK, learning assumptions, WACC) to scenarios that explicitly represent diffusion limits and competition for
new-build slots, rather than assuming unconstrained adoption. 

A second constraint emphasized in the note is that even if fusion becomes cost-competitive, \emph{deployment rates themselves are likely
S-curve limited} by industrial scale-up, supply chains, siting/permitting throughput, and financing capacity (Figure~2) \cite{Cardozo2019Valley}.
This framing is consistent with the broader ``valley of death'' argument: large, long-lead investments and construction durations can suppress
the innovation cycle and slow diffusion even when long-run economics look attractive. For the CATF IWG methodology, the practical implication
is that deployment modeling should be treated as a distinct analysis layer that (i) translates annual build-rate ceilings into cumulative
capacity trajectories, (ii) maps those trajectories onto learning-by-doing assumptions (i.e., learning depends on realized production volume),
and (iii) allows scenario stress-tests where the cost-out path is decoupled from, or constrained by, build-rate limits.

\section{Worked Examples and Open-Source Colab Workflow}
A key objective of the CATF IWG development was that the methodology extensions not remain ``paper-only,'' but be
\emph{executable, reproducible, and auditable} by external users. Accordingly, the CATF work was implemented and
exercised through a live Python workflow that couples a python package (pyFECONs) to template-driven \LaTeX{} report generation (automatic population of tables, cost-account rollups, and assumption documentation). This approach ensures that each published result can be traced to intermediate computed quantities (e.g., power-balance terms, geometry-driven volumes/areas, driver electrical requirements), mapped into the
standards-aligned chart-of-accounts, and then rolled up deterministically into capital cost categories and baseline LCOE.  This capability is made accessible through a web portal \cite{CATF_FusionCostModelPortal}. 

A separate colab notebook contains the optional analysis layers developed under CATF---probabilistic costing (materials,
TRL, and learning compounding), safety-informed scoping calculations with cost-account mapping, and extended
economic measures beyond LCOE---so that users can reproduce both baseline and extended outputs from the same
COA-mapped plant model. 

The worked examples in the colab are therefore intended as \emph{entry points} into the open workflow rather than as stand-alone derivations. Each example corresponds to a self-contained notebook section that (i) states the engineering assumptions and cost bases, (ii) computes the relevant sizing or proxy quantities (e.g., first-wall/blanket/shield
demonstrations; driver-centric account development in 22.1.3 and power-supply refinements in 22.1.7; safety-driven
provisions such as bioshield thickness, detritiation proxies, or site-boundary implications), and (iii) writes the resulting
values into the report templates so that the example outputs appear in consistent COA tables and narratives. For
transparency and reuse, the complete CATF IWG implementation notebook is publicly accessible as a Google Colab
reference implementation \cite{CATFColab2025}, and the ARPA-E baseline methodology and COA lineage that it
extends are documented in the companion ARPA-E support report \cite{Woodruff2026CostingFramework}.

\section{Discussion: Relationship to Prior Power-Plant and Fusion Costing Methodologies}
\label{sec:discussion_prior_methodologies}

Fusion costing has historically progressed along two partially independent tracks. One track produced
\emph{concept-specific} integrated studies---often exemplary in technical depth for a chosen architecture, but difficult to
compare across concepts because accounting structures, cost boundaries, and embedded assumptions differed (e.g., Sheffield’s
Generomak/ESECOM-era systems models for magnetic fusion; LIFE for inertial fusion). A second track, developed primarily for
fission and other large energy infrastructure, emphasized \emph{standardized accounting and reporting} through
internationally recognized charts of accounts (IAEA/GIF-EMWG/EPRI) and associated estimating conventions. The ARPA-E fusion
costing capability (2017--2024), and the CATF IWG extensions (2024--2025), intentionally synthesize these tracks: the
approach retains the \emph{standards-aligned COA as the stable comparability container} while embedding within that container
a physics-to-plant workflow and auditable subsystem realizations that can be applied consistently across \emph{MFE, IFE, and
MIFE}. The practical consequence is a methodology that is less tied to any single design point than earlier concept studies,
and more focused on producing traceable, updateable, and portfolio-consistent estimates in which dominant fusion-unique cost
drivers are explicitly modeled rather than implicitly embedded.

The key contrast with concept-specific fusion costing is therefore not that the underlying engineering philosophy differs
(indeed, both traditions recognize that economics only make sense when physics, engineering constraints, and plant
integration are internally consistent), but that ARPA-E+CATF formalize \emph{(i) a common accounting spine} and \emph{(ii) a
controlled set of ``swap-in'' driver models} so that differing architectures can be compared without rewriting the entire
economic framework. In ARPA-E+CATF, the COA does the work of standardization, while methodological refinement is concentrated
in those accounts where fusion uncertainty and cost leverage are largest---notably the dominant driver and its closely-coupled
electrical infrastructure (Accounts~22.1.3 and~22.1.7), with safety-informed cost bases propagated into other impacted
accounts and then rolled up transparently into total capital cost and LCOE.

\subsection{Standards-aligned COA as a common spine: beyond concept-specific accounting}
The GIF Economics Modeling Working Group (EMWG) guidelines and the IAEA/EPRI COA lineage were developed to enable consistent
reporting and comparison of complex nuclear plants by prescribing a harmonized accounting structure, reporting conventions,
and clear distinctions between top-down and bottom-up estimate types \cite{gif-emwg-2007,iaea-2001,epri-2024}. In this
tradition, the COA is not a cost model by itself; it is the \emph{accounting container} that makes cost models comparable and
auditable.

Earlier fusion studies often did not operate inside such a standardized container. Sheffield’s Generomak/ESECOM-era models
were explicitly ``generic'' and comparative within magnetic fusion, but their cost categories and scalings were not designed
for direct alignment with regulated power-plant COAs \cite{sheffield-generomak,sheffield-milora-2016}. LIFE and related IFE
studies developed detailed subsystem models tailored to inertial fusion, but likewise employed concept-specific cost
structures that are not automatically interoperable with other fusion architectures or fission COA reporting
\cite{life-ife}. By contrast, ARPA-E+CATF adopt the COA explicitly as a compatibility layer: results can be reported in the
same major accounts as conventional plants while still allowing fusion-unique subsystems to be treated with the fidelity
needed for credible comparisons. This design choice is central to portfolio decision-making: it reduces ambiguity about what
is included in ``capital cost,'' improves interpretability by non-specialists, and enables future updates without
re-baselining the entire methodology each time a subsystem model is improved.

\subsection{Driver swap-in philosophy: magnets (MFE), lasers (IFE), pulsed power (MIFE)}
A second defining difference from concept-specific studies is the deliberate separation between a \emph{stable common plant
framework} and a \emph{small set of architecture-defining driver models}. In many prior analyses, the driver technology and
its integration assumptions are inseparable from the overall costing structure: the accounting, scalings, and even the
economic framing are tuned to the concept. ARPA-E+CATF instead treat the dominant driver account as a controlled swap-point.
In the standards-aligned COA, Account~22.1.3 (``Coils or Lasers or Pulsed Power'') is used as the consistent location for the
dominant fusion-unique driver, while Account~22.1.7 (``Power Supplies'') captures the coupled electrical infrastructure. Under
CATF, these accounts are the primary locus of refinement: magnets (TF/PF/CS) are swapped in for MFE, lasers/driver modules
for IFE, and pulsed-power systems for MIFE, each with explicit cost bases, scaling logic, and (where relevant) lifetime and
replacement treatments. This structure enables learning, TRL/maturity uncertainty, and supply-chain-informed updates to be
applied directly to the most consequential components without losing COA comparability.

This driver swap-in philosophy provides an explicit mechanism for what concept-specific studies achieve implicitly: it keeps
the integrated physics-to-plant workflow (the strength of ARIES/LIFE-like studies), but constrains where architectural
differences enter the accounting so that cross-architecture comparisons do not conflate bookkeeping differences with
technology differences. In this sense, ARPA-E+CATF can be read as a methodological bridge between ``deep'' concept studies and
the needs of a multi-architecture, rapidly evolving private-sector portfolio.

\subsection{Relationship to ARIES-style systems studies: common workflow, different objective function}
ARIES and related fusion systems studies established the durable backbone for fusion plant evaluation: couple a physics
performance model to an engineering-constrained radial build, translate to subsystem sizing, and estimate costs using a mix
of bottoms-up elements and scaling relations. ARPA-E+CATF retain this backbone, but the objective function differs. Classic
ARIES studies generally aim to optimize and report a coherent point design (or small set of variants) within a given
architecture class. ARPA-E+CATF must instead support \emph{portfolio-consistent evaluation} across heterogeneous architectures
and maturity levels, which shifts the emphasis toward: (i) standardized accounting boundaries, (ii) explicit indirect cost
and EPC/constructability sensitivity, and (iii) auditable traceability and updateability of cost bases as concepts evolve.

In practice, this produces a different balance between concept-specific detail and reusable structure: ARPA-E+CATF preserve a
common COA and conventional BOP anchoring so that changes in the fusion island propagate consistently into plant-level
results, while concentrating detailed modeling effort in the dominant cost drivers (22.1.3 and 22.1.7) and in the safety-
affected accounts that drive licensing and site boundary implications. This keeps the integrated workflow of ARIES but
reduces the degree to which each concept requires a bespoke costing ``dialect.''

\subsection{Contrast with Sheffield/Generomak and LIFE: concept-specific integration versus standardized comparability}
Sheffield’s Generomak/ESECOM-era tradition is a foundational antecedent: it recognized early that economic conclusions depend
on internally consistent physics/engineering/economics, and it provided a reusable framework for comparing magnetic fusion
variants \cite{sheffield-generomak}. LIFE and related IFE studies similarly developed integrated models that treated the
driver and plant integration explicitly and produced coherent cost and performance scaling relationships for inertial fusion
\cite{life-ife}. These approaches remain valuable as deep technical references and as sources of subsystem scaling insight.

The ARPA-E+CATF methodology differs primarily in \emph{structural intent}. Rather than producing a best-estimate cost model
optimized around a particular concept, ARPA-E+CATF aim to provide a \emph{standards-aligned, auditable evaluation framework}
that can ingest multiple concepts with minimal redefinition of accounting boundaries. This structural choice forces explicit
treatment of what is often implicit in concept studies: the separation of driver costs from supporting electrical plant
(22.1.3 vs 22.1.7), the explicit roll-up of indirect costs and execution pathways, and the capacity to propagate safety basis
assumptions into affected accounts. The result is not a substitute for concept-specific engineering studies; it is a
portfolio-grade costing infrastructure that can incorporate such studies as improved cost bases while preserving a stable
comparability layer.

\subsection{Summary of distinguishing characteristics}
In summary, ARPA-E+CATF are best understood as a modernization and operationalization of prior fusion costing traditions for
a multi-architecture, rapidly evolving context:
\begin{itemize}
  \item They \emph{inherit} the integrated physics-to-plant ethos of ARIES and earlier systems-code traditions (including
  Generomak/ESECOM and LIFE), but separate that workflow from concept-specific accounting.
  \item They \emph{adopt} the IAEA/GIF/EPRI COA lineage as a stable reporting and comparability spine, improving alignment
  with conventional power-plant costing practice.
  \item They \emph{extend} prior practice by formalizing a driver swap-in structure (22.1.3) coupled to explicit electrical
  infrastructure treatment (22.1.7), and by emphasizing auditable cost bases, indirect-cost transparency, and safety-informed
  propagation into other accounts, ending in comparable capital cost and LCOE outputs.
\end{itemize}
These choices reflect the thesis that fusion economic credibility, especially in portfolio settings, requires not only
integrated engineering, but also standardized accounting and traceability: dominant cost drivers must be explicit,
comparable, and updateable as subsystem assumptions evolve and as concepts progress toward pilot-plant design.

\subsection{Forward-looking: the next phase of fusion power-plant costing and analysis}
\label{subsec:future_costing}

Fusion power-plant costing is rapidly transitioning from document-centric, point-design reporting toward
\emph{data-rich, optimization-driven, and provenance-complete} digital workflows in which economics is evaluated
concurrently with geometry and physics/engineering feasibility. A clear emerging pattern is the coupling of integrated
multi-physics plant models to multi-objective optimization loops, with increasing reliance on surrogate modeling to make
exploration of high-dimensional design spaces computationally tractable and to enable publication of Pareto trade spaces
rather than single ``best'' designs \cite{DigitalFusion2024}. Recent work by nTtau Digital exemplifies
this direction by embedding costing directly into automated co-design: parametric CAD geometry generation (plasma boundary,
coils, structures, and plant integration) feeds multi-physics analyses (electromagnetics, structural mechanics, thermal
analysis, and optional neutronics), and \emph{as each component is instantiated}, extracted quantities (mass, volume,
surface area, power ratings, and complexity measures) are mapped into a standardized code-of-accounts (COA). The result is
an auditable ``cost build file'' per design point that aggregates component estimates into subsystem costs and plant-level
totals, with provenance capture (inputs, versions, solver settings, and outputs) and scalable orchestration across compute
resources \cite{nTtauDigital2026Workflow}. In parallel, broader practice is converging on integrated whole-plant
environments that evaluate plasma, magnets, structures, blankets, power conversion, and \emph{cost} in a unified workflow,
then generate Pareto-optimal sets using evolutionary   multi-objective optimizers \cite{DigitalFusion2024}.

Equally important, the next phase is not only about deeper optimization, but also about \emph{accessibility and reuse} of
the underlying costing and economics models through web-based interfaces. Under CATF, a key emphasis is to keep the
pyFECONs costing standard intact while lowering the barrier to use: the same standards-aligned COA and workflow can
be exposed through a web portal that guides users through valid input sets, runs the model in a controlled environment,
and returns COA roll-ups, capital costs, and LCOE outputs with traceable assumptions (mirroring the open Python/Colab
workflow used for the CATF worked examples) \cite{CATFColab2025}. Complementarily, Fusion Advisory Services is developing a
web portal for the economics modeling extensions---so that outputs from the costing backbone (CAPEX/OPEX, availability,
financing assumptions, learning and uncertainty settings) can be transformed into investment and planning measures (e.g.,
revenue requirements, NPV/IRR/MIRR, and scenario-based deployment implications) in a user-accessible, reproducible format.
This ``web-first'' accessibility theme is also reflected in Hugo Bellows Weeks' open-source software framing for fusion
market and economic analysis, to be presented at the upcoming OSSFE workshop \cite{Weeks2026ForecastingFusionMarketplace}.
Taken together, these trends point toward a tooling ecosystem where (i) standards-aligned accounting remains the durable
interface for comparability, (ii) optimization-driven design workflows continuously generate COA-mapped cost builds, and
(iii) web-accessible front-ends make both costing and downstream economics/deployment analyses usable by a broader
community without sacrificing traceability or methodological discipline \cite{nTtauDigital2026Workflow,DigitalFusion2024}.

\section{Further work}
\label{sec:further_work}

The costing framework described here is intended to be extensible: its value increases as additional physics, engineering,
safety, and supply-chain realism are incorporated while preserving the same standards-aligned chart-of-accounts (COA)
mapping and auditability. Over the next year, the CATF IWG will shift from ad hoc development to a \emph{monthly release and
community-engagement cadence} focused on maturing the dominant driver account (22.1.3) across MFE/IFE/MIFE, tightening the
supporting electrical infrastructure account (22.1.7), and expanding the shared cost-basis library, governance, and
traceability expectations for broad stakeholder adoption \cite{CATFcolab}. This cadence is designed not only
to improve the technical fidelity of the model, but also to accelerate convergence on a common economic ``language'' for
fusion: if universally adopted by developers, EPCs, suppliers, and funders, a standards-aligned COA can enable clearer
procurement signals, faster diligence, and more direct identification of supply-chain gaps and scale-up priorities.

\subsection{Monthly roadmap: driver account maturation (22.1.3) and coupled electrical infrastructure (22.1.7)}
The monthly working group meetings are organized as three architecture-specific development threads---MFE, IFE, and
MIFE---with periodic cross-cutting sessions on cost bases and indirect accounts. The near-term plan is to deliver and
stress-test ``v1'' driver modules for each architecture in Account~22.1.3 (magnets for MFE; lasers for IFE; pulsed power for
MIFE), followed by successive refinements that add learning, manufacturing breakdowns, installation and integration
fidelity, and explicit lifetime/replacement logic \cite{CATFcolab}. In parallel, each driver thread will
tighten the boundary and interface to Account~22.1.7 (Power Supplies) so that supporting electrical plant costs are
represented consistently across architectures (e.g., wall-plug power implications for IFE, charging and pulse-forming
infrastructure for MIFE, and conventional electrical support for MFE) and are not obscured inside single-factor driver
scalings \cite{CATFcolab}. Each monthly session includes a ``model clinic'' to ensure the code remains
accessible through the open Colab/web workflow and to accelerate incorporation of community feedback via issues and pull
requests \cite{CATFcolab}.

\subsubsection{Planned implementation of AACE estimate accuracy ranges}

To improve consistency and transparency in future CATF IWG cost products, we plan to align
reported cost uncertainty with the \emph{AACE International Cost Estimate Classification System}.
In this framework, the \emph{accuracy range} is the interval within which there is an 80\% confidence
that the actual cost will fall, often described as the project ``cone of uncertainty'' as definition
matures from early screening to detailed design.

Practically, this will be implemented as follows. First, each CATF estimate deliverable will be
assigned an AACE estimate class based on the maturity of the underlying project definition
(e.g., availability of design basis, equipment lists, quantities, vendor quotes, and schedule logic).
Second, reported cost results will include an uncertainty band consistent with the corresponding
AACE class (Table~\ref{tab:aace_accuracy_ranges}). Third, contingency and risk adjustments will be
reported in a manner that is traceable to (i) estimate class, (ii) key drivers of uncertainty (technical,
regulatory, supply-chain, schedule), and (iii) any probabilistic analysis used to support the 80\%
confidence interpretation.

\begin{table}[htbp]
\centering
\caption{AACE International cost estimate classification and typical accuracy ranges (80\% confidence).}
\label{tab:aace_accuracy_ranges}
\begin{tabular}{llll}
\hline
\textbf{Estimate Class} & \textbf{Maturity Level} & \textbf{Typical Purpose} & \textbf{Expected Accuracy Range (Low/High)} \\
\hline
Class 5 & 0\% to 2\%   & Screening or Feasibility & $-30\%$ to $-50\%$ \; / \; $+30\%$ to $+100\%^{*}$ \\
Class 4 & 1\% to 15\%  & Concept Screening        & $-15\%$ to $-30\%$ \; / \; $+20\%$ to $+50\%$ \\
Class 3 & 10\% to 40\% & Budget Authorization     & $-10\%$ to $-20\%$ \; / \; $+10\%$ to $+30\%$ \\
Class 2 & 30\% to 75\% & Control or Bid           & $-5\%$ to $-15\%$  \; / \; $+5\%$ to $+20\%$ \\
Class 1 & 65\% to 100\%& Check Estimate           & $-3\%$ to $-10\%$  \; / \; $+3\%$ to $+15\%$ \\
\hline
\end{tabular}

\vspace{2mm}
\footnotesize{$^{*}$Class 5 ranges are often highly asymmetric and can be wider depending on novelty,
scope ambiguity, and data limitations.}
\end{table}

\subsection{Safety-informed costing as a design-coupled cost driver}
A priority extension is tighter integration of fusion safety analysis into the costing workflow, informed by the ongoing
CATF IWG safety thread embedded across the monthly sessions. The methodological objective is to treat safety not as a
post-hoc checklist but as a \emph{design-coupled cost driver}: confinement strategy, hazard controls (cryogens, high voltage,
laser hazards), tritium inventory limits, waste classification approach, and operational constraints each impose tangible
implications for facility layout, ventilation and detritiation capacity, shielding, remote maintenance provisions,
commissioning scope, and staffing. In practice, the next-year plan is to formalize a small set of \emph{safety basis input
knobs} that (i) add or modify specific COA accounts (buildings/structures, auxiliary systems, radioactive waste treatment,
instrumentation and control, and owner's costs), and (ii) propagate into indirect accounts through QA/QC requirements,
testing/acceptance, licensing scope, and commissioning complexity \cite{CATFcolab}. This will enable more
consistent sensitivity studies of the cost consequences of safety posture choices across MFE/IFE/MIFE.

\subsection{Cost-basis library, governance, and supply-chain alignment to the COA}
Several of the monthly sessions explicitly target ``cost bases'' as a first-class community product: a curated, versioned
library of cost basis entries with metadata (source type, normalization to installed cost, reference year dollars, and
uncertainty tags) and clear contribution templates \cite{CATFcolab}. This work is directly aligned with the
broader goal of accelerating deployment through \emph{shared standards}. If stakeholders converge on COA-based
categorization, the supply chain can be described in a way that is immediately interpretable across developers and
reviewers: vendors can map products and services into cost accounts; EPCs can bid, benchmark, and de-risk projects using a
common accounting structure; and funding organizations can identify capability gaps (manufacturing capacity, qualifying
test infrastructure, long-lead items, workforce needs) in an apples-to-apples way across concepts. A concrete near-term
example is the ongoing effort to categorize fusion supply chains according to standardized cost accounts (including work by
Fusion Advisory Services), which can translate ``what is missing'' from qualitative narratives into account-tagged
procurement and scale-up priorities that are legible to both industry and capital providers. The IWG will use the monthly
forum to refine this taxonomy, establish governance expectations for public cost bases, and encourage EPC-facing
interpretation and validation of account boundaries and indirect-cost treatment \cite{CATFcolab}.

\subsection{Power-flow visualization and traceability outputs}
A tractable improvement planned for the next cycle is richer visualization of the internal power accounting and its
relationship to cost drivers. The current workflow computes the major terms in the plant power balance (fusion power
partition, thermal conversion, recirculating loads, and net electric output), but results are typically reported as tables.
Embedding Sankey-diagram generation as a first-class output would improve interpretability for non-specialists, strengthen
internal consistency checks (e.g., avoiding double counting between thermal and direct conversion channels), and provide a
clearer bridge between recirculating power fractions and the sizing/cost of subsystems that create those loads (notably
drivers and power supplies). This is particularly useful when comparing fuel cycles and direct energy conversion options,
where the qualitative flow structure changes.

\subsection{Time-dependent analysis}

Schwartz et al.\ embed fusion power plants in a capacity-expansion model of a deeply decarbonized U.S.\ grid to quantify the system value of fusion as a function of cost, availability, and operating flexibility, rather than relying only on plant-level LCOE \cite{Schwartz2023ValueFusionJoule}. They show that fusion’s optimal deployment and resulting value are highly sensitive to achievable overnight capital cost and reliability, with significant adoption only when fusion reaches competitive cost and performance targets within stringent emissions-constrained scenarios.

\subsection{Materials cost bases and the challenge of FOAK--NOAK transitions}
Finally, the monthly materials thread will focus on improving transparent cost bases for fusion-relevant materials and
purity specifications. While NOAK costing is essential for long-term comparisons, credible NOAK unit costs are not readily
available in open sources for several fusion-relevant materials at required purity/form factors. The next-year plan is to
combine (i) vendor engagement and anonymized price curve development, (ii) explicit specification of purity and form
factor, (iii) yield/scrap/recycle treatment, and (iv) scenario analysis that separates FOAK procurement conditions from
NOAK steady-state production \cite{CATFcolab}. In the model, this translates into improved unit cost inputs,
manufacturing multipliers reflecting realistic fabrication routes, and uncertainty bounds for materials-dominated accounts
that can be propagated through total capital cost and LCOE.

\subsection{Summary}
In sum, the next phase of CATF IWG work is organized around a practical monthly roadmap: mature the driver ``swap-in''
modules in Account~22.1.3 for MFE/IFE/MIFE, strengthen the coupled electrical infrastructure representation in Account~22.1.7,
and institutionalize a shared, governed cost-basis library with explicit safety, materials, and indirect-cost threads
\cite{CATFcolab}. Beyond improving model fidelity, these steps aim to accelerate fusion deployment by making
costing and supply-chain discussions legible across stakeholders: a universally adopted standards-aligned COA can provide a
common procurement and diligence language for developers, EPCs, suppliers, and funders, enabling faster identification of
resource gaps, clearer investment signals, and more coordinated industrialization pathways toward commercial fusion.

\section{Conclusion}
Fusion is entering a phase in which economic credibility, comparability, and traceability are as consequential as physics
performance. As concepts mature from laboratory demonstrations toward integrated pilot plants and early commercial
facilities, stakeholders increasingly require cost estimates that are (i) transparent and auditable, (ii) comparable across
diverse architectures, and (iii) traceable to explicit technical assumptions rather than embedded in opaque scaling
relations or single-point levelized outputs. This paper documents how the CATF International Working Group (IWG) on
Fusion Cost Analysis (2024--2025) advanced that requirement by extending---without replacing---the ARPA-E
standards-aligned fusion costing backbone (2017--2024) implemented in pyFECONs. The central organizing principle is
to retain the internationally recognizable chart-of-accounts (COA) and physics-to-economics workflow as the stable
interface for comparability, while increasing fidelity precisely where fusion cost uncertainty is highest: the dominant
fusion-unique drivers and their coupled electrical infrastructure, plus safety- and regulation-linked plant provisions.
In particular, the CATF IWG reorganized development around three architecture-defining tracks (MFE, IFE, and MIFE)
and treated Account~22.1.3 as a controlled ``swap-point'' that is replaced by an explicit cost-account development for the
dominant driver in each class---magnets (TF/PF/CS) for MFE, lasers/driver modules for IFE, and pulsed power for MIFE---with
Account~22.1.7 refined in parallel to represent supporting power supplies and pulse-forming infrastructure using
auditable installed-cost bases and lifetime/replacement logic. These changes preserve like-for-like COA roll-ups to total
capital cost and baseline LCOE while enabling more defensible cross-architecture comparisons grounded in requirements,
geometry, and explicit cost bases.

Beyond driver cost realism, the CATF IWG implemented modular extensions that convert a deterministic costing backbone
into an extensible analysis environment. First, probabilistic costing was introduced to compound three dominant sources
of uncertainty---materials price dispersion, TRL-based maturity uncertainty, and learning-rate uncertainty---into cost
distributions that remain attributable to specific accounts rather than being treated as external adders. The IWG also
outlined a planned alignment of reported uncertainty bands with the AACE estimate-class accuracy ranges (the ``cone of
uncertainty''), providing a disciplined mechanism to communicate estimate maturity consistently across deliverables.
Second, a safety-informed costing module was developed that enumerates fusion-relevant hazards, translates them into
mitigating systems/structures/provisions, and maps these explicitly into the affected COA accounts, supplemented by
scenario-parameterized regulatory and financial adders (licensing fees and insurance). Third, the economics layer was
expanded beyond LCOE to include value and planning measures (NPV, IRR/MIRR, revenue requirements, WACC-based
annualization, and residual/follow-on value), computed directly from the same COA-mapped outputs; and the deployment
modeling extension was positioned as a complementary layer that bounds plausible scale-up trajectories using diffusion
and build-rate constraints and couples these to learning-by-doing assumptions. Collectively, these additions strengthen
the original ARPA-E thesis: fusion economic credibility requires not only better physics and better components, but
better accounting---explicit, comparable, and updateable as subsystem assumptions evolve. 

Finally, this work underscores a broader implication for accelerating fusion deployment: the long-run value of these tools
depends on stakeholder convergence around shared standards. A universally adopted, standards-aligned COA provides a
common language for developers, EPCs, suppliers, regulators, and funders to interpret scope, benchmark costs, and
identify missing capabilities. When dominant subsystems and supporting infrastructure are consistently categorized by
account (e.g., driver hardware in 22.1.3, power-conditioning infrastructure in 22.1.7, safety-driven provisions mapped
into buildings, fuel-cycle, radioactive systems, and financial adders), supply-chain capacity gaps and investment needs
can be expressed in procurement-ready terms rather than qualitative narratives. In that framing, the CATF IWG
extensions should be viewed as enabling infrastructure: a standards-preserving, driver-centric, uncertainty- and
safety-aware costing environment that supports credible pilot-plant and NOAK scenario analysis, provides traceable
inputs for finance and deployment modeling, and offers a pathway to web-accessible, reproducible engagement without
sacrificing auditability or comparability.

\section*{Acknowledgements}

This work was performed with support from the Clean Air Task Force across the 2024-2025 period, including continued evolution of the toolchain toward a closed-source branch with a web interface and expanded costing features. 

We acknowledge the CATF IWG on Fusion Cost Analysis, including Laila El-Guebaly and Geoffrey Rothwell for guidance and constructive feedback on costing structure, reporting conventions, and economic interpretation.  We acknowledge deep discussions with Patrick White, who leads the CATF IWG on Safety, on the taxonomy of safety hazards and the methods for computation of the costs associated with their amelioration.

\bibliographystyle{unsrtnat}
\bibliography{catfreferences}

\end{document}